\documentclass[pdftex,singlecolumn,epjc3]{svjour3} 
\RequirePackage{graphicx}
\RequirePackage{mathptmx}      
\RequirePackage{flushend}
\RequirePackage[numbers,sort&compress]{natbib}
\RequirePackage[colorlinks,citecolor=blue,urlcolor=blue,linkcolor=blue]{hyperref}
\usepackage{amsmath}
\usepackage{graphicx}
\usepackage{geometry}
\usepackage{caption}
\usepackage{subcaption}
\usepackage{color}
\begin{document}
\baselineskip=18 pt
\begin{center}
{\large{\bf Topological Effects on Non-Relativistic Eigenvalue Solutions Under AB-Flux Field with Pseudoharmonic- and Mie-type Potentials  }}
\end{center}

\vspace{0.1cm}

\begin{center}
{\bf Faizuddin Ahmed}\footnote{\bf faizuddinahmed15@gmail.com ; faizuddin@ustm.ac.in}\\
{\bf Department of Physics, University of Science \& Technology Meghalaya,\\
Ri-Bhoi, Meghalaya-793101, India }
\end{center}

\vspace{0.1cm}

\begin{abstract}:
In this paper, we investigate the quantum dynamics of a non-relativistic particle confined by the Aharonov-Bohm quantum flux field with pseudoharmonic-type potential in the background of topological defect produced by a point-like global monopole. We solve the radial Schr\"{o}dinger equation analytically and determine the exact eigenvalue solution of the quantum system. Afterwards, we consider a Mie-type potential in the quantum system and solve the radial equation analytically and obtain the eigenvalue solution. We analyze the effects of the topological defect and the quantum flux with these potentials on the energy eigenvalue and wave function of the non-relativistic particles. In fact, it is shown that the energy levels and wave functions are influenced by the topological defect shifted the result compared to the flat space results. In addition, the quantum flux field also shifted the eigenvalue solutions and an analogue of the Aharonov-Bohm effect for bound-states is observed. Finally, we utilize these eigenvalue solutions to some known diatomic molecular potential models and presented the energy eigenvalue and wave function.  
\end{abstract}

\vspace{0.5cm}


\vspace{0.5cm}

{\bf PACS Number(s):} 14.80.Hv, 03.65.Ge, 02.30.Gp, 03.65.Vf, 03.65.-w

\section{Introduction}
\label{intro}

The searching for the exact or approximate eigenvalue solutions of the non-relativistic and relativistic wave equations have turned to be a principal part from the beginning of quantum mechanics \cite{bb1}. These exact and approximate solutions of the wave equations are important in different fields, such as atomic physics, nuclear and high energy physics, quantum electrodynamics, and theory of molecular vibrations \cite{bb2,bb5,bb8,bb9,bb11,bb12,bb14,bb15,bb16}. The exact or approximate eigenvalue solutions of the stationary Schr\"{o}dinger wave equation plays an important role to examine the correctness of models and approximations in computational physics as well as in chemistry. In addition, particle interacts with potential of various kinds have used to describe importance of several physical systems since they contain all the necessary information of a quantum state under investigation. However, analytic solutions are possible only for a few simple quantum systems such as the hydrogen atom and harmonic oscillator problem that were given in many textbooks \cite{bb1,bb17,bb18,bb19,bb20,bb21}. To obtain the exact and approximate solutions of the wave equations, various methods or techniques have employed, such as the Nikiforov–Uvarov method \cite{bb22,bb23,bb24,bb25,bb26,bb27} and its functional analysis, supersymmetric quantum mechanics approach (SUSYQM) \cite{bb28,bb29}, asymptotic iteration method (AIM) \cite{bb30,bb31}, Laplace's transformation method, and special functions of various kinds. 

Furthermore, studies of the quantum motions of charged particles in the presence of an uniform magnetic and the Aharonov-Bohm (AB) flux fields, which are perpendicular to the plane where the particles are confined have been done since past few years. In fact, the study of the non-relativistic and relativistic charged particles that are confined to the magnetic field has been a growing research interested because of possible applications in different fields, such as in graphene \cite{bb15,bb16,bb33}, semiconductor structures \cite{bb36}, chemical physics \cite{bb38}, biology \cite{bb39}, molecular vibrational and rotational spectroscopy in molecular physics \cite{bb40}. The investigation of the particles with magnetic field have been done by many authors in the literature including two-dimensional solid with Kratzer potential in the presence of screw dislocation \cite{ME5} and in cosmic string space-time \cite{bb41,bb411,bb412,bb413}. Note that the presence of the Aharonov-Bohm flux field in quantum mechanical systems shows an analogue of the Aharonov-Bohm (AB) effect \cite{YA,MP}. This AB-effect is a quantum mechanical phenomenon in which an electrically charged particle is affected by the quantum flux field despite being confined to a region in which both the magnetic and electric fields are zero. The experimental confirmation of its existence was presented in Ref. \cite{NO}. 

Another kind of system that has been studied in the context of quantum mechanical problems are the topological defects which were formed in the early universe through symmetry-breaking mechanism \cite{TWBK}. Topological defects are classified into cosmic strings \cite{SZ2}, domain walls, and global monopoles \cite{MBAV,ERBM,ERBM2}. The quantum mechanical systems in the background of a space-time generated by a cosmic string have been studied by several authors (see, Ref. \cite{bb41,bb411,bb412,bb413,ME} and related references therein). The spherically symmetric static point-like global monopole space-time was first presented in Ref. \cite{MBAV} and have been studied both in the relativistic limit (see, Refs. \cite{ALCO2,ALCO,SR}) and only a handful works in the non-relativistic limit \cite{CF,CF2,VBB,PN,FA} in quantum system. The presence of the topological defect in a quantum system changes its physical properties and modifies the energy eigenvalue and wave function of a particle compared to the flat space result. 

Our aim in this work is to study the quantum motions of the non-relativistic particles under the influences of the Aharonov-Bohm flux field in the background of a point-like global monopole with different interactions potential, such as pseudoharmonic-type and Mie-type potentials. In fact, we determine the exact eigenvalue solutions analytically and analyze the effects of various factors. Afterwards, we use these eigenvalue solutions for some known molecular potential models and show the effects of the topological defect and the quantum flux field which modified the results compared to the flat space results obtained in the literature.

This paper is organized as follows: in {\tt section 2}, we discuss the three-dimensional radial Schr\"{o}dinger wave equation in the background of a point-like global monopole under the flux field with potential. Then, we solve this radial equation with two types of potential, namely a pseudoharmonic-type potential ({\tt sub-section 2.1}) and the Mie-type potential ({\tt sub-section 2.2}) and obtain the eigenvalue solutions. In {\tt section 3}, we use these eigenvalue solutions to some some known potential models in the literature; and in {\tt section 4}, we present our results. Throughout the analysis, we use the natural units ${\rm c=1=\hbar}$.

\section{Topological Effects on Radial Solution Under AB-flux Field with Potential}

A static and spherically symmetric point-like global monopole space-time in the spherical coordinates $({\rm t, r, \phi, \theta})$ is given by \cite{ERBM2,CF,CF2,ALCO,SR,FA,PN}
\begin{equation}
{\rm ds}^2=-{\rm dt}^2+\frac{{\rm dr}^2}{\alpha^2}+{\rm r}^2\,({\rm d\theta}^2+{\rm \sin}^2 \theta\,{\rm d\phi}^2)=-{\rm dt}^2+{\rm g_{ij}}\,{\rm dx^{i}}\,{\rm dx^{j}},
\label{1}
\end{equation}
where $\alpha<1$ represents the topological defect parameter of a point-like global monopole (PGM), ${\rm g_{ij}}$ is the spatial metric tensor whose nonzero components are ${\rm g_{rr}=\frac{1}{\alpha^2}}$, ${\rm g_{\theta\theta}=r^2}$, ${\rm g_{\phi\phi}=r^2\,\sin^2 \theta}$ with their contravariant components ${\rm g^{rr}=\alpha^2}$, ${\rm g^{\theta\theta}=\frac{1}{r^2}}$, ${\rm g^{\phi\phi}=\frac{1}{r^2\,\sin^2 \theta}}$. This point-like global monopole space-time have many interesting features including a naked curvature singularity on the symmetry axis ${\rm R=R^{\,\mu}_{\mu}=\frac{2\,(1-\alpha^2)}{r^2}}$. Other properties of this geometry are given in Refs. \cite{ERBM2,ALCO}.

In this section, we study the quantum motions of the non-relativistic particle under a potential in the presence of the Aharonov-Bohm flux field in this point-like global monopole background. Afterwards, we will consider the Mie-type potential and obtain the eigenvalue solution analytically and discuss the effects of topological defects as well as quantum flux on the eigenvalue solutions. Therefore, the time-dependent Schr\"{o}dinger wave equation in the presence of an electromagnetic potential $\vec{A}$ is described by the following wave equation \cite{CF,CF2,PN,FA,ME2,ME5}
\begin{eqnarray}
{\rm \Big[-\frac{1}{2\,M}\,\Big(\frac{1}{\sqrt{g}}\,D_{i}\,(\sqrt{g}\,g^{ij}\,D_{j})\Big)+V(r)\Big]\,\Psi=i\,\frac{d\Psi}{dt}},
\label{2}
\end{eqnarray}
where ${\rm M}$ is the rest mass the particles, ${\rm g}$ is the determinant of the metric tensor ${\rm g_{ij}}$ with ${\rm g^{ij}}$ its inverse, and ${\rm D_{i}=(\partial_{i}-i\,e\,A_{i})}$ with ${\rm e}$ is the electric charges. For the space-time (\ref{1}) under consideration, the determinant of the spatial metric tensor ${\rm g_{ij}}$ is given by ${\rm g=\frac{r^2\,\sin^2\,\theta}{\alpha^2}}$.

By the method of separation of variables, one can write the total wave function ${\rm \Psi (t, r, \theta, \phi)}$ in terms of different variables. Suppose, a possible wave function in terms of a radial wave function ${\rm \psi (r)}$ is as follows:
\begin{equation}
{\rm \Psi(t, r, \theta, \phi)=e^{-i\,E\,t}\,Y_{l,m} (\theta, \phi)\,\psi(r)},
\label{3}
\end{equation}
where ${\rm E}$ is the non-relativistic particles energy, ${\rm Y_{l,m} (\theta, \phi)}$ is the spherical harmonic functions, and ${\rm l, m}$ are respectively the orbital and magnetic moment quantum numbers.

In this analysis, we chosen the electromagnetic three-vector potential $\vec{A}$ given by Refs. \cite{ALCO,SR,FA}
\begin{equation}
{\rm A_{r}=0=A_{\theta}},\quad {\rm A_{\phi}=\frac{\Phi_B}{2\,\pi\,r\,\sin \theta}},\quad {\rm \Phi_B=\Phi\,\Phi_0},\quad {\rm \Phi_0=2\,\pi\,e^{-1}},
\label{4}
\end{equation}
where ${\rm \Phi_B}=const$ is the Aharonov-Bohm flux, ${\rm \Phi_0}$ is the quantum of magnetic flux, and ${\rm \Phi}$ is the amount of magnetic flux which is a positive integer.

Thereby, expressing the wave equation (\ref{2}) in the space-time background (\ref{1}) and using Eqs. (\ref{3})--(\ref{4}), we obtain the following radial and angular differential equations for ${\rm \psi (r)}$ and ${\rm Y_{l,m}}$, respectively as
\begin{equation}
{\rm \frac{\alpha^2}{r^2}\,\frac{d}{dr}\,\left(r^2\,\frac{d\psi (r)}{dr}\right) +\left[2\,M\,\Big(E-V(r)\Big)-\frac{l'\,(l'+1)}{r^2}\right]\,\psi(r)=0},
\label{5}
\end{equation}
And
\begin{equation}
{\rm \left[\frac{1}{\sin\theta}\,\frac{d}{d\theta}\,\left(\sin\theta\, \frac{d}{d\theta}\right)+\frac{1}{\sin^2\theta}\,\left(\frac{d}{d\phi}-i\,\Phi\right)^2+l'\,(l'+1)\right]\,Y_{l,m} (\theta,\phi)=0}
\label{5a}
\end{equation}
where ${\rm l'=(l-\Phi)}$ is the effective orbital quantum number.
 
The radial Schr\"{o}dinger equation can be written as 
\begin{eqnarray}
&&{\rm \psi''(r)+\frac{2}{r}\,\psi'(r)+\frac{1}{\alpha^2}\,\Bigg[-\frac{l'\,(l'+1)}{r^2}+2\,M\,\Big(E-V(r)\Big)\Bigg]\,\psi(r)=0}.
\label{6}
\end{eqnarray}
The effective potential of the system (by changing the function ${\rm \psi(r)=\frac{R(r)}{r}}$ in the Eq. (\ref{6}), one can find the one-dimensional Schrodinger-like equation) given by
\begin{equation}
{\rm V_{eff} (r)=\frac{1}{\alpha^2}\,\Bigg[V(r)+\frac{(l-\Phi)\,(l-\Phi+1)}{2\,M\,r^2}\Bigg]}.
\label{7}
\end{equation}
We can see that the effective potential of the quantum system depends on the topological defect characterized by the parameter ${\rm \alpha}$, and the quantum flux field ${\rm \Phi_{AB}}$.

In this analysis, we consider the following two different kinds of potential of physical interest and determines the exact eigenvalue solutions of the non-relativistic wave equation analytically.

\subsection{\bf Effects of Pseudoharmonic-type Potential}

\begin{figure}
\begin{subfigure}[b]{0.4\textwidth}
\includegraphics[width=3.0in,height=1.5in]{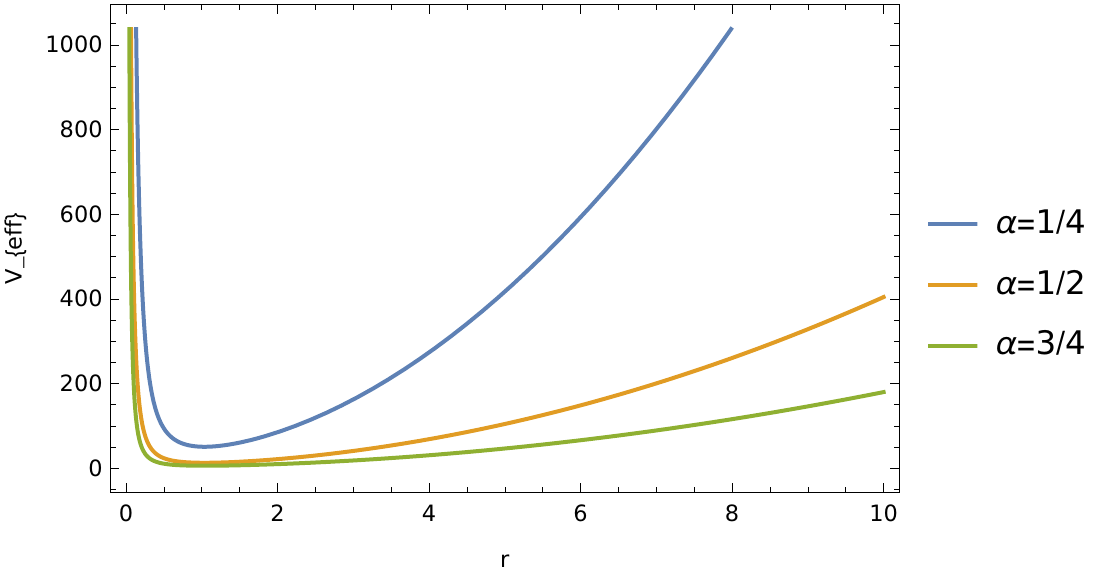}
\caption{$l=1=M=\beta=\gamma=V_0$, $\Phi=3/4$.}
\label{fig: 1 (a)}
\end{subfigure}
\hfill
\begin{subfigure}[b]{0.4\textwidth}
\includegraphics[width=2.8in,height=1.5in]{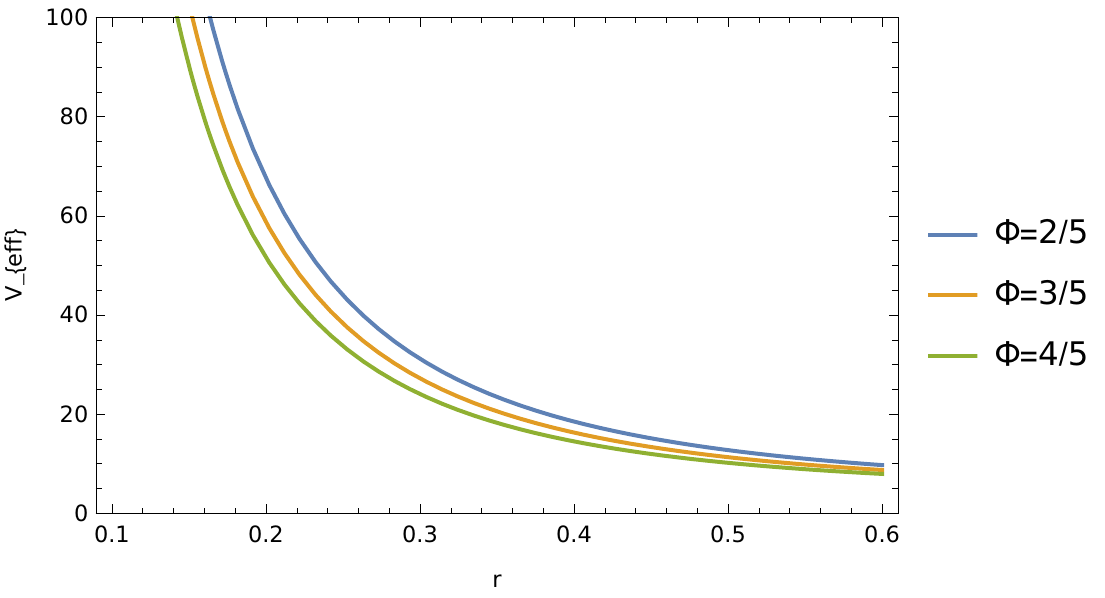}
\caption{$l=1=M=\beta=\gamma=V_0$, $\alpha=3/4$.}
\label{fig: 1 (b)}
\end{subfigure}
\hfill\\
\begin{subfigure}[b]{0.4\textwidth}
\quad\includegraphics[width=3.0in,height=1.5in]{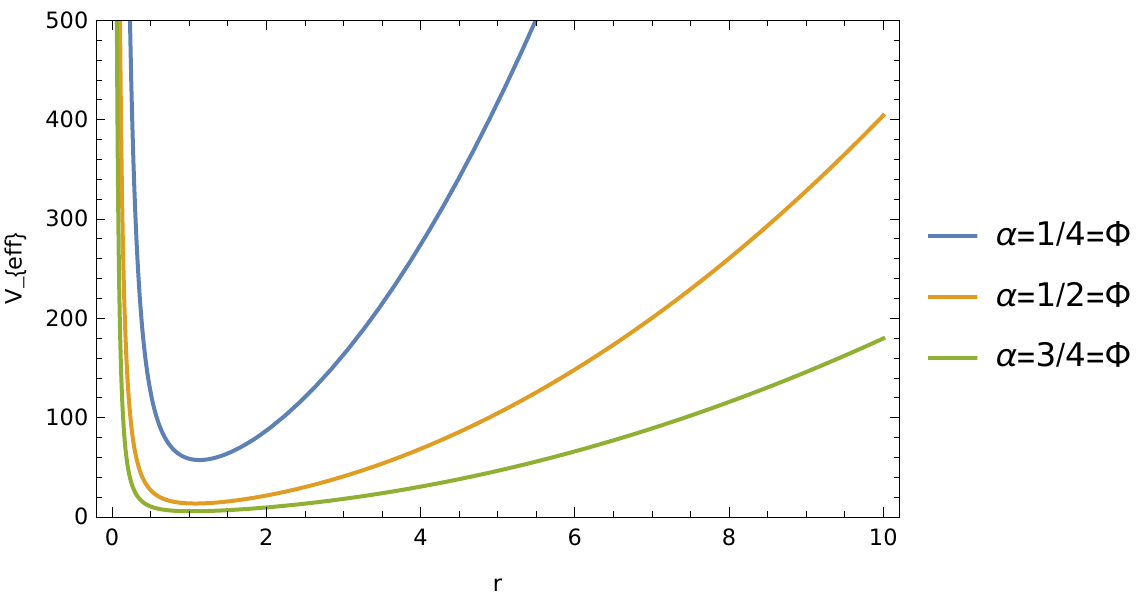}
\caption{$l=1=M=\beta=\gamma=V_0$.}
\label{fig: 1 (c)}
\end{subfigure}
\hfill
\begin{subfigure}[b]{0.4\textwidth}
\includegraphics[width=2.6in,height=1.5in]{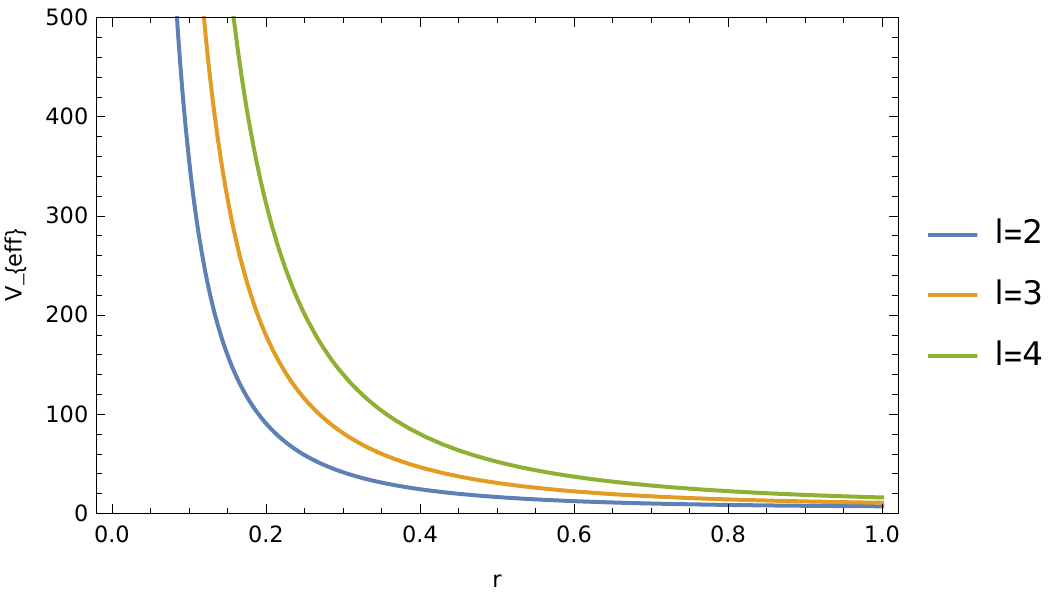}
\caption{$l=1=M=\beta=\gamma=V_0$, $\alpha=3/4$, $\Phi=1$.}
\label{fig: 1 (d)}
\end{subfigure}
\caption{The effective potential of the system with $r$ for values of the topological defect parameter, and the magnetic flux with quantum number.}
\label{fig: 1}
\end{figure}

In this section, we are interested on the following potential $V (r)$ superposition of harmonic oscillator and inverse quadratic potential given by \cite{SMI,SHD,KJO,SMI3,VK} 
\begin{equation}
{\rm V(r)=V_0+\beta\,r^2+\frac{\gamma}{r^2}},
\label{8}
\end{equation}
where ${\rm \beta, \gamma}$ characterise the potential parameters and ${\rm V_0>0}$ is a constant potential term.

Thereby, substituting potential (\ref{8}) in the Eq. (\ref{6}), we obtain the following radial equation:
\begin{equation}
{\rm \psi''(r)+\frac{2}{r}\,\psi'(r)+\Big(\Lambda-\frac{j^2}{r^2}-\omega^2\,r^2\Big)\,\psi(r)=0},
\label{9}
\end{equation}
where we set the parameters
\begin{eqnarray}
{\rm \Lambda=\frac{2\,M\,(E-V_0)}{\alpha^2}}\quad,\quad
{\rm j=\sqrt{\frac{(l-\Phi)\,(l-\Phi+1)+2\,M\,\gamma}{\alpha^2}}}\quad,\quad {\rm \omega=\sqrt{\frac{2\,M\,\beta}{\alpha^2}}}.
\label{10}
\end{eqnarray}
The effective potential of the quantum system using potential (\ref{8}) will be given by
\begin{equation}
{\rm V_{eff} (r)=\frac{1}{\alpha^2}\,\Bigg[V_0+\beta\,r^2+\frac{\gamma}{r^2}+\frac{(l-\Phi)\,(l-\Phi+1)}{2\,M\,r^2}\Bigg]}.
\label{11}
\end{equation}
We see that the effective potential under consideration depends on the topological defect, the background curvature as well as the magnetic flux. We have plotted few graphs (fig. 1) showing the influences of various factors on the effective potential.

Transforming the above equation (\ref{9}) via ${\rm \psi (r)=\frac{U (r)}{r^{3/2}}}$, we obtain the following equation
\begin{equation}
{\rm U''(r)-\frac{1}{r}\,U'(r)+\Bigg[\Lambda-\frac{(j^2-\frac{3}{4})}{r^2}-\omega^2\,r^2\Bigg]\,U(r)=0}.
\label{12}
\end{equation}
Introducing a new variables via ${\rm s=\omega\,r^2}$ in the above Eq. (\ref{12}), we obtain the following second-order differential equation:
\begin{equation}
{\rm U''(s)+\left(\frac{1-4\,\mu^2}{4\,s^2}\right)\,U(s)+\frac{\nu}{s}\,U(s)-\frac{1}{4}\,U(s)=0},
\label{13}
\end{equation}
where different parameters are defined as
\begin{equation}
{\rm 2\,\mu=\sqrt{j^2+\frac{1}{4}}}\quad,\quad {\rm \nu=\frac{\Lambda}{4\,\omega}}.
\label{14}
\end{equation}
Equation (\ref{13}) is the Whittaker differential equation \cite{KDM} and ${\rm U(s)}$ is the Whittaker function which we can write in terms of the confluent hypergeometric function of the first kind ${\rm {}_1 F_{1} (s)}$ \cite{KDM,MA,AP,LJS,GBA} as follows
\begin{equation}
{\rm U (s)=s^{\frac{1}{2}+\mu}\,e^{-\frac{s}{2}}\,{}_1 F_{1}\left(\mu-\nu +\frac{1}{2},2\,\mu+1; s\right)}.
\label{15}
\end{equation}

Our aim is for searching the bound-state solutions of the quantum system by solving the equation (\ref{13}). It is well-known that the confluent hypergeometric function ${}_1 F_{1}(s)$ becomes a finite degree polynomial in $s$ of degree ${\rm n}$ by imposing that ${\rm \Big(\mu-\nu+\frac{1}{2}\Big)}$ is a negative integer, that is, ${\rm \Big(\mu-\nu+\frac{1}{2}\Big)=-n}$, where ${\rm n=0,1,2,..}$. 

After simplification of the condition ${\rm \Big(\mu-\nu+\frac{1}{2}\Big)=-n}$ using Eq. (\ref{14}) and then Eq. (\ref{10}), one will find the following expression of the energy eigenvalue ${\rm E_{n,l}}$ given by
\begin{eqnarray}
{\rm E_{n,l}=V_0+\alpha\,\sqrt{\frac{2\,\beta}{M}}\,\Bigg(2\,n+\sqrt{\frac{(l-\Phi)\,(l-\Phi+1)+2\,M\,\gamma}{\alpha^2}+\frac{1}{4}}+1\Bigg)}.
\label{16}
\end{eqnarray}
The normalized radial wave functions are given by
\begin{equation}
{\rm \psi_{n,l} (s)=D_{n,l}\,\Bigg(\frac{2\,M\,\beta}{\alpha^2}\Bigg)^{3/8}\,s^{\mu-\frac{1}{4}}\,e^{-\frac{s}{2}}\,{}_1 F_{1}(-n,2\,\mu+1 ; s)},
\label{17}
\end{equation}
where ${\rm D}_{n,l}$ is a constant that can be determined by the normalization condition for the radial wave function \cite{EAFB}
\begin{equation}
{\rm \frac{1}{\alpha}\,\int^{\infty}_{0}\,r^2\,dr\,|\psi(r)|^2=1}.
\label{18}
\end{equation}
To solve integral of the radial wave function, one can write the confluent hypergeometric function in terms of the associated Laguerre polynomials given by the relation \cite{GBA}
\begin{equation}
{\rm {}_1 F_{1}\left(-n,2\,\mu+1; x\right)=\frac{n!\,(2\,\mu)!}{(n+2\,\mu)!}\,L^{(2\,\mu)}_{n} (x)}.
\label{19}
\end{equation}
Then, taking into account ${\rm s=\omega\,r^2}$, and with the help of \cite{AP} to solve the integrals, the normalization constant is given by
\begin{equation}
{\rm D_{n,l}=\frac{1}{(2\,\mu)!}\,\sqrt{\frac{2\,\alpha\,(n+2\,\mu)!}{n!}}=\frac{1}{\left(\sqrt{j^2+\frac{1}{4}}\right)!}\,\sqrt{\frac{2\,\alpha\,\left(n+\sqrt{j^2+\frac{1}{4}}\right)!}{n!}}}.
\label{20}
\end{equation}

\begin{figure}
\begin{subfigure}[b]{0.4\textwidth}
\includegraphics[width=2.8in, height=1.45in]{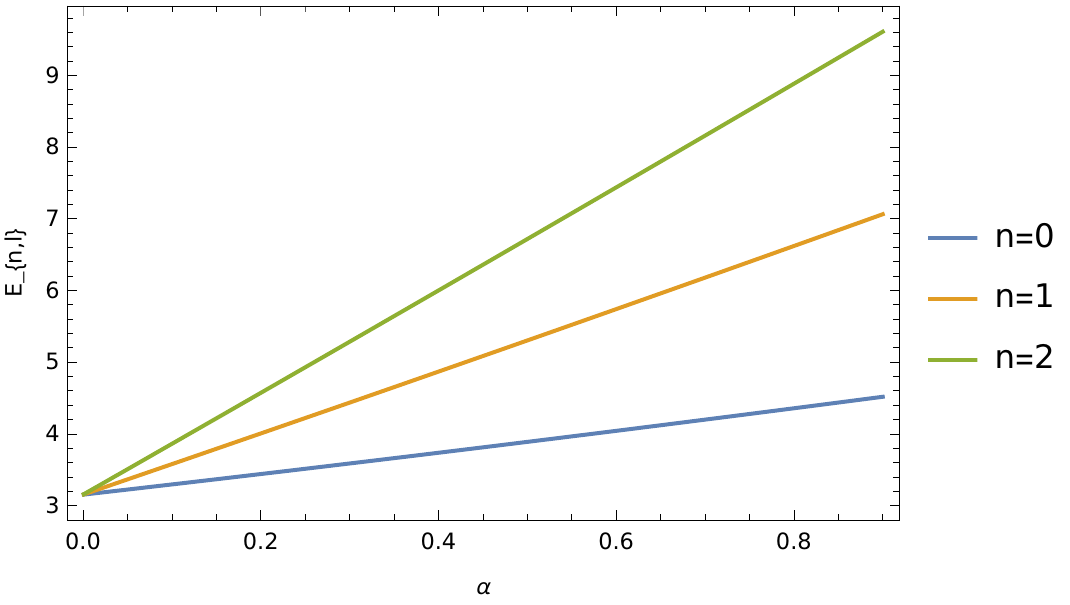}
\caption{$l=1=M=\beta=\gamma=V_0$, $\Phi=3/4$.}
\label{fig: 2 (a)}
\end{subfigure}
\hfill
\begin{subfigure}[b]{0.4\textwidth}
\includegraphics[width=3.1in, height=1.45in]{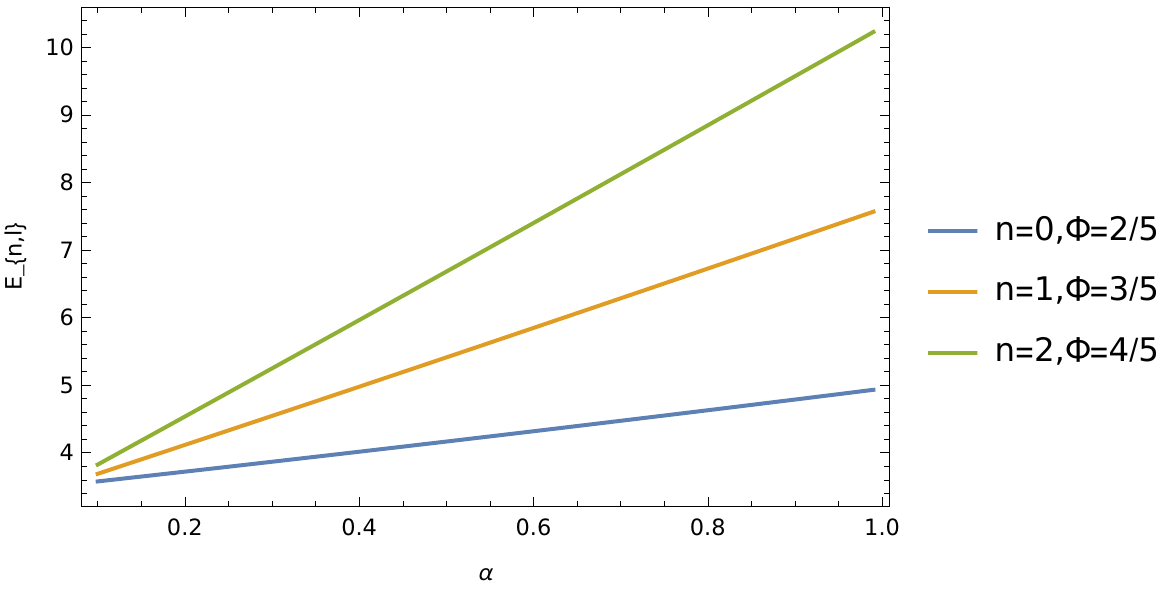}
\caption{$l=1=M=\beta=\gamma=V_0$.}
\label{fig: 2 (b)}
\end{subfigure}
\hfill\\
\begin{subfigure}[b]{0.4\textwidth}
\includegraphics[width=2.9in, height=1.45in]{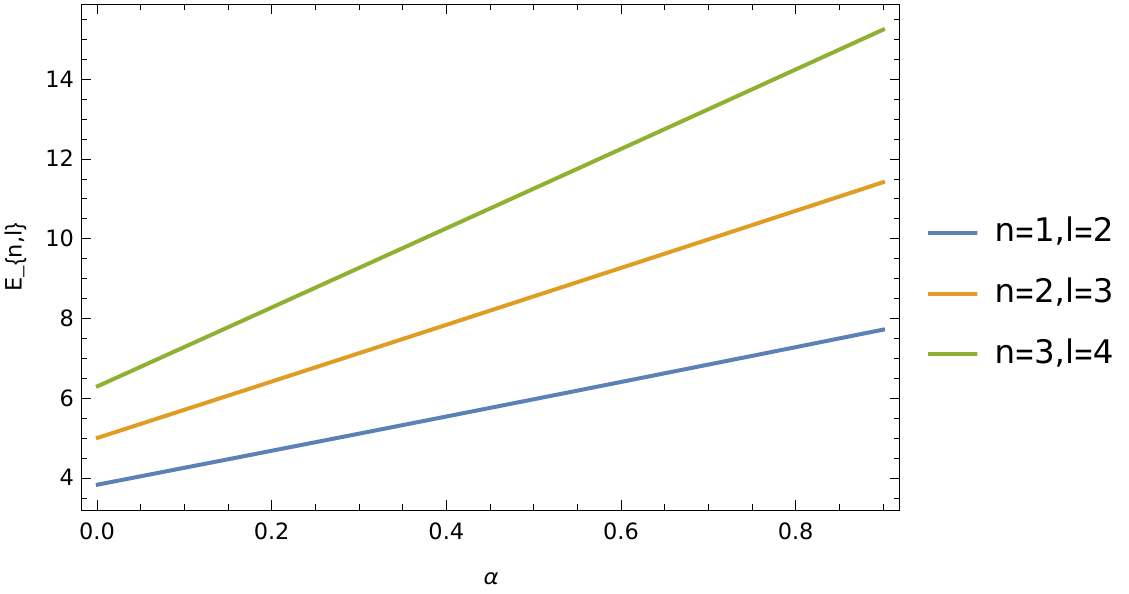}
\caption{$M=1=\beta=\gamma=V_0=\Phi$.}
\label{fig: 2 (c)}
\end{subfigure}
\hfill
\begin{subfigure}[b]{0.4\textwidth}
\includegraphics[width=2.8in, height=1.45in]{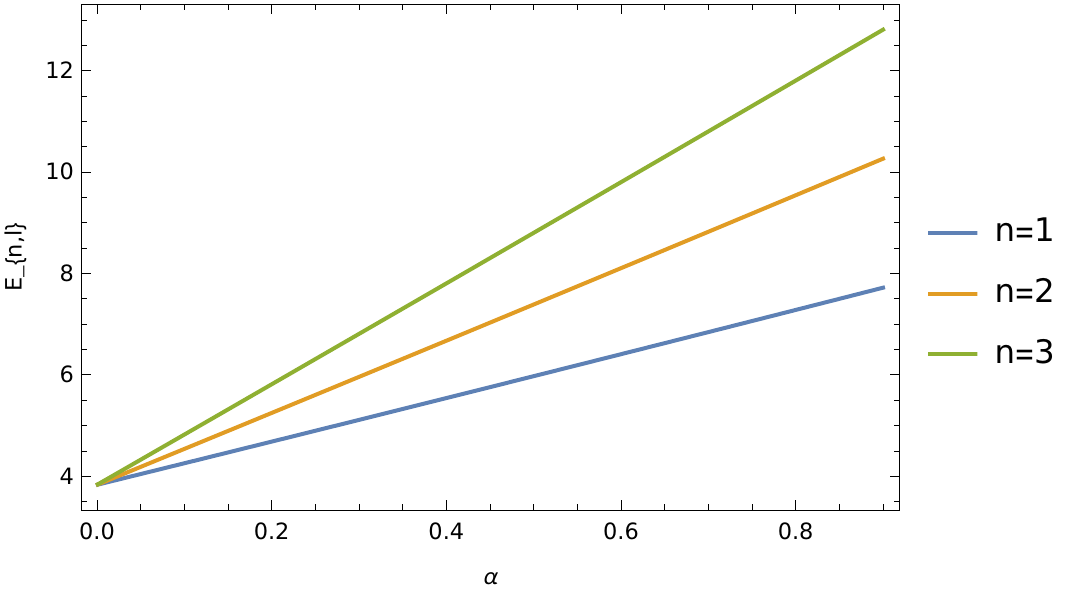}
\caption{$l=1=M=\beta=\gamma=V_0$, $\Phi=0$.}
\label{fig: 2 (d)}
\end{subfigure}
\caption{The energy eigenvalue $E_{n,l}$ with topological defect parameter $\alpha$ for different values of other parameters.}
\label{fig: 2}
\end{figure}

Equation (\ref{16}) is the energy levels and Eqs. (\ref{17})--(\ref{20}) are the normalized radial wave functions of a non-relativistic particle confined by the AB-flux field with pseudoharmonic-type potential under the topological defects produced by a point-like global monopole. One can see that the energy levels and the radial wave functions are influenced by the topological defect characterized by the parameter ${\rm \alpha}$, and the quantum flux field ${\rm \Phi_{AB}}$ and get them modified compared to the flat space results obtained in Refs. \cite{SMI,SHD,KJO,SMI3,VK}. We have plotted a few graphs of the energy levels ${\rm E_{n,l}}$ (fig. 2) and the normalized radial wave function ${\rm \psi_{n,l}(s)}$ (fig. 3) for different values of various parameters. 

Below, few special cases of the above quantum system will be discussed. 

\begin{figure}
\begin{subfigure}[b]{0.4\textwidth}
\includegraphics[width=3.0in, height=1.45in]{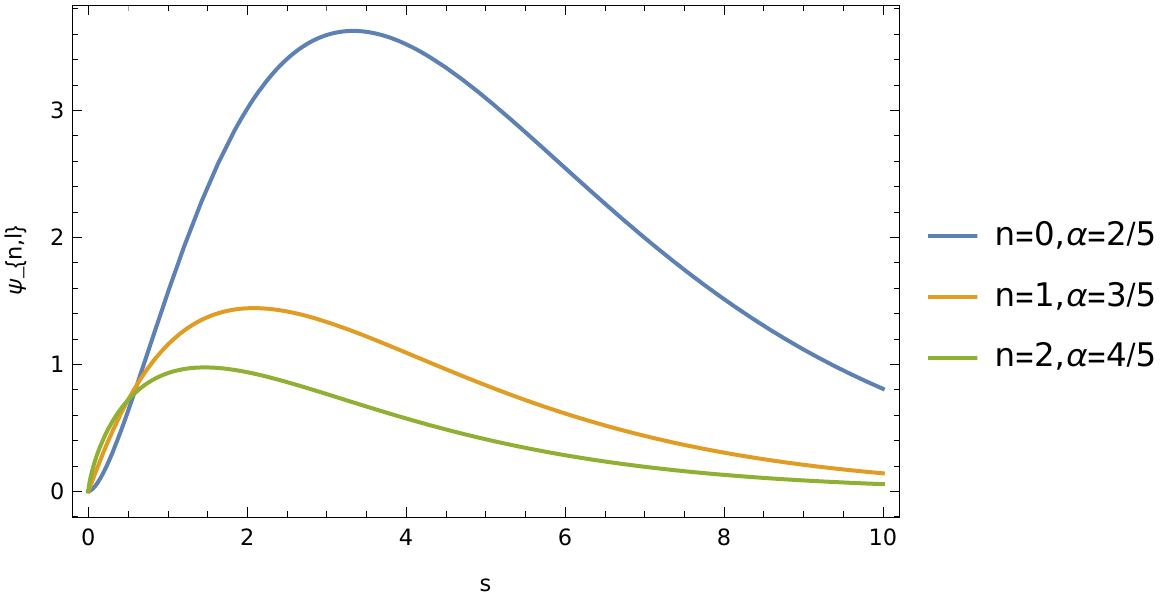}
\caption{$l=1=M=\beta=\gamma$, $\Phi=3/4$.}
\label{fig: 3 (a)}
\end{subfigure}
\hfill
\begin{subfigure}[b]{0.4\textwidth}
\includegraphics[width=3.0in, height=1.45in]{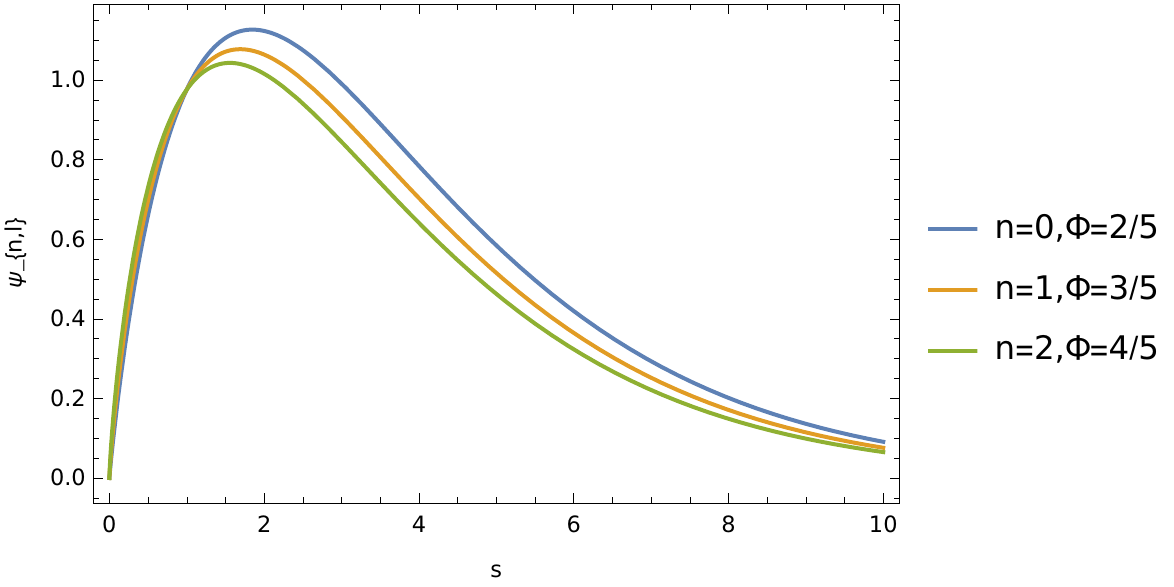}
\caption{$l=1=M=\beta=\gamma$, $\alpha=3/4$.}
\label{fig: 3 (b)}
\end{subfigure}
\hfill\\
\begin{subfigure}[b]{0.4\textwidth}
\includegraphics[width=3.1in, height=1.45in]{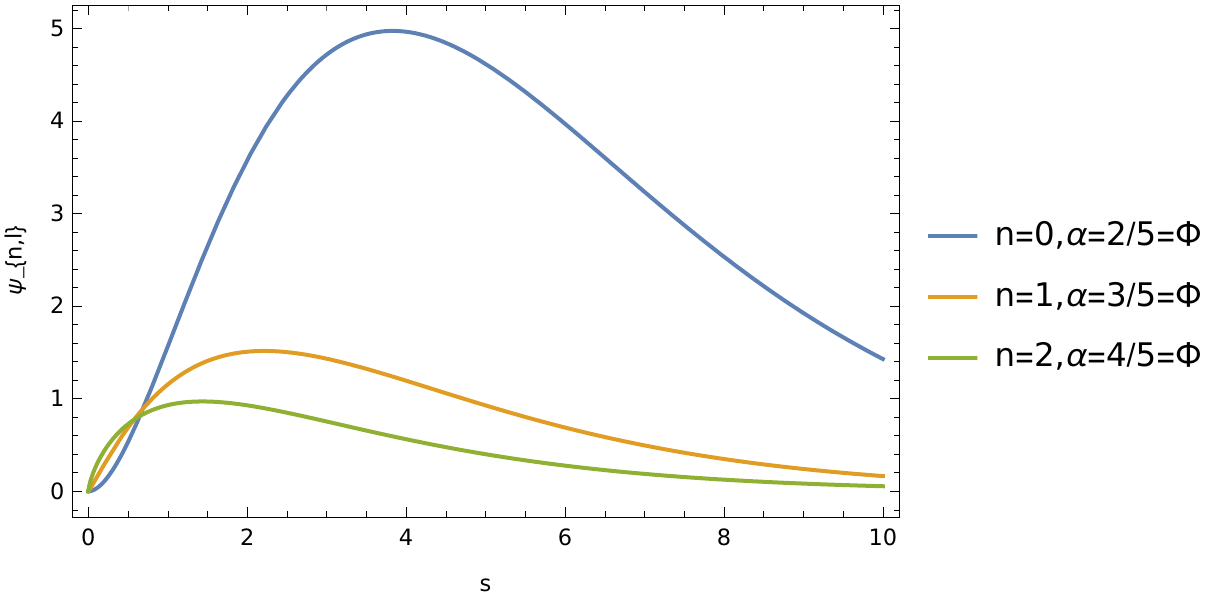}
\caption{$l=1=M=\beta=\gamma$.}
\label{fig: 3 (c)}
\end{subfigure}
\hfill
\begin{subfigure}[b]{0.4\textwidth}
\includegraphics[width=2.9in, height=1.45in]{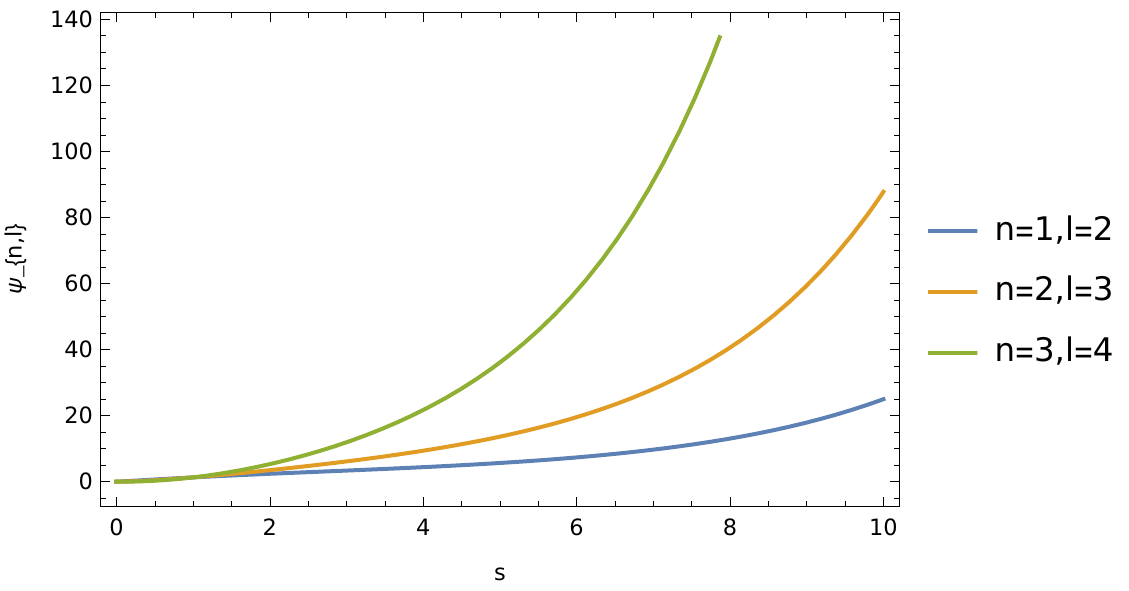}
\caption{$M=1=\beta=\gamma$, $\alpha=3/4$, $\Phi=1$.}
\label{fig: 3 (d)}
\end{subfigure}
\caption{The normalized radial wave function $\psi_{n,l}$ for different values of the parameters.}
\label{fig: 3}
\end{figure}

\vspace{0.3cm}
{\bf Case I: Without Topological defect and Potential ${\rm V (r)=\Big(\frac{A}{r^2}+B\,r^2\Big)}$.}
\vspace{0.3cm}

In this case, we want to study the above quantum system without topological defects. In that case, for $\alpha \to 1$, the space-time geometry (\ref{1}) under consideration becomes Minkowski flat space. Furthermore, we set the constant potential term $V_0=0$ in the potential expression (\ref{8}).  

Therefore, the energy eigenvalue in that case from (\ref{16}) becomes
\begin{eqnarray}
{\rm E_{n,l}=\sqrt{\frac{2\,B}{M}}\,\Bigg(2\,n+1+\sqrt{\Big(l-\Phi+\frac{1}{2}\Big)^2+2\,M\,A}\Bigg)}.
\label{21a}
\end{eqnarray}
which is similar to the energy eigenvalue obtained in Ref. \cite{FC} provided zero magnetic flux ${\rm \Phi \to 0}$ here. Thus, the magnetic flux considered in the quantum system shifts the eigenvalue solution of a non-relativistic particle in comparison to the known result obtained in Ref. \cite{FC} with potential of the form ${\rm V(r)=\Big(\frac{A}{r^2}+B\,r^2\Big)}$. Overall we can see that the energy eigenvalue (\ref{16}) gets more modified in comparison to the result in Ref. \cite{FC} due to the presence of the topological defects characterise by the parameter $\alpha$, and the constant potential term $V_0$ in addition to the magnetic flux field ${\rm \Phi_{AB}}$ in the quantum system.

The normalized radial wave functions in that case becomes
\begin{equation}
{\rm \psi_{n,l}(s)=N_{n,l}\,\Bigg(\frac{2\,M\,B}{\alpha^2}\Bigg)^{3/8}\,s^{\bar{\mu}-\frac{1}{4}}\,e^{-\frac{s}{2}}\,{}_1 F_{1}\left(-n,1+2\,\bar{\mu}; s\right)},
\label{21b}
\end{equation}
where 
\begin{eqnarray}
{\rm N_{n,l}}={\rm \frac{1}{\left(\sqrt{\vartheta^2+\frac{1}{4}}\right)!}\,\sqrt{\frac{2\,\left(n+\sqrt{\vartheta^2+\frac{1}{4}}\right)!}{n!}}},\quad {\rm \vartheta=\sqrt{(l-\Phi)\,(l-\Phi+1)+2\,M\,A}},\quad 2\,\bar{\mu}=\sqrt{\vartheta^2+\frac{1}{4}}.
\label{21c}
\end{eqnarray}

\vspace{0.3cm}
{\bf Case II: Without Topological Defects but Potential ${\rm V(r)=\Big(\frac{A}{r^2}+B\,r^2+C\Big)}$}
\vspace{0.3cm}

We discuss the above quantum mechanical system without topological defects, that is, $\alpha \to 1$. In that case, the space-time geometry (\ref{1}) under consideration reduces to Minkowski flat space. In addition, the chosen potential here is of the form ${\rm V(r)=\Big(\frac{A}{r^2}+B\,r^2+C\Big)}$.    

Therefore, for $\alpha \to 1$ and with the chosen potential form, one will have the following energy eigenvalue expression
\begin{eqnarray}
{\rm E_{n,l}=C+\sqrt{\frac{2\,B}{M}}\,\Bigg(2\,n+1+\sqrt{\Big(l-\Phi+\frac{1}{2}\Big)^2+2\,M\,A}\Bigg)}
\label{21d}
\end{eqnarray}
which is similar to the result obtained in the flat space Ref. \cite{bb40} provided zero magnetic flux ${\rm \Phi \to 0}$ here. Thus, the magnetic flux considered in the quantum system shifts the eigenvalue solution of a non-relativistic particle compared to the known result in \cite{bb40} with this chosen potential. Hence, overall we can see that the topological defects of a point-like global monopole characterize by the parameter $\alpha$, and the quantum flux field ${\rm \Phi_{AB}}$ modified the eigenvalue solution compared to the result obtained in \cite{bb40} with this potential. The normalized radial wave function will be the same obtained in Eqs. (\ref{21b})--(\ref{21c}) with now ${\rm \bar{\Lambda}=2\,M\,(E_{n,l}-C)}$.

\subsection{\bf Effects of Mie-Type Potential}

In this section, we are interested on another kind of potential equal to repulsive Coulomb plus inverse square potential given by \cite{SMI,KJO,SMI3,SMI2,SMI4,ME3,ME4}
\begin{equation}
{\rm V(r)=V_0+\frac{\delta}{r}+\frac{\gamma}{r^2}},
\label{22}
\end{equation}
where ${\rm \delta, \gamma}$ characterizes the potential parameter. 

For this type of potential, the effective potential of the system becomes 
\begin{equation}
{\rm V_{eff} (r)=\frac{1}{\alpha^2}\,\Bigg[V_0+\frac{\delta}{r}+\frac{\gamma}{r^2}+\frac{(l-\Phi)\,(l-\Phi+1)}{2\,M\,r^2}\Bigg]}.
\label{22a}
\end{equation}
We can see that the effective potential of the system depends on the topological defect ans well as the magnetic flux. We have plotted a few graph (fig. 4) showing the influences of various parameters on the effective potential.

\begin{figure}
\begin{subfigure}[b]{0.4\textwidth}
\includegraphics[width=2.8in, height=1.45in]{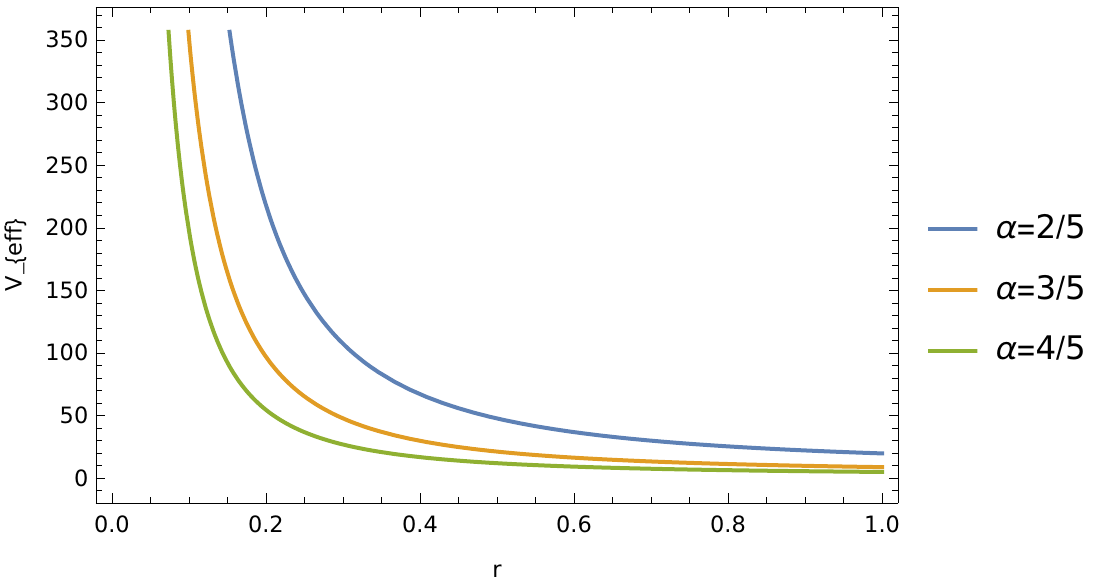}
\caption{$l=1=M=\delta=\gamma=V_0$, $\Phi=3/4$.}
\label{fig: 4 (a)}
\end{subfigure}
\hfill
\begin{subfigure}[b]{0.4\textwidth}
\includegraphics[width=2.8in, height=1.45in]{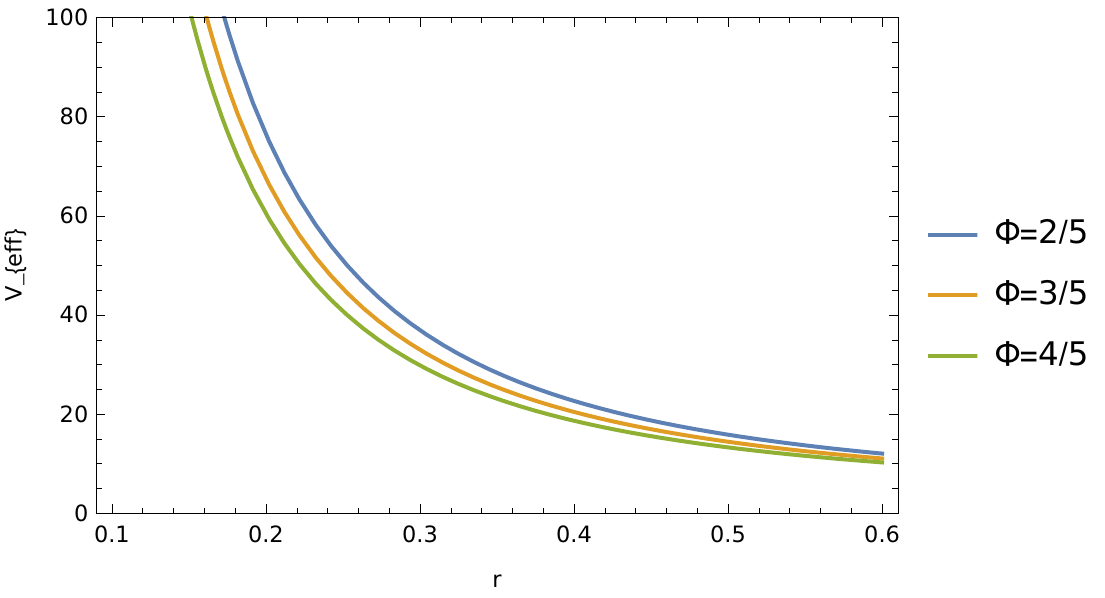}
\caption{$l=1=M=\delta=\gamma=V_0$, $\alpha=3/4$.}
\label{fig: 4 (b)}
\end{subfigure}
\hfill\\
\begin{subfigure}[b]{0.4\textwidth}
\includegraphics[width=2.9in, height=1.45in]{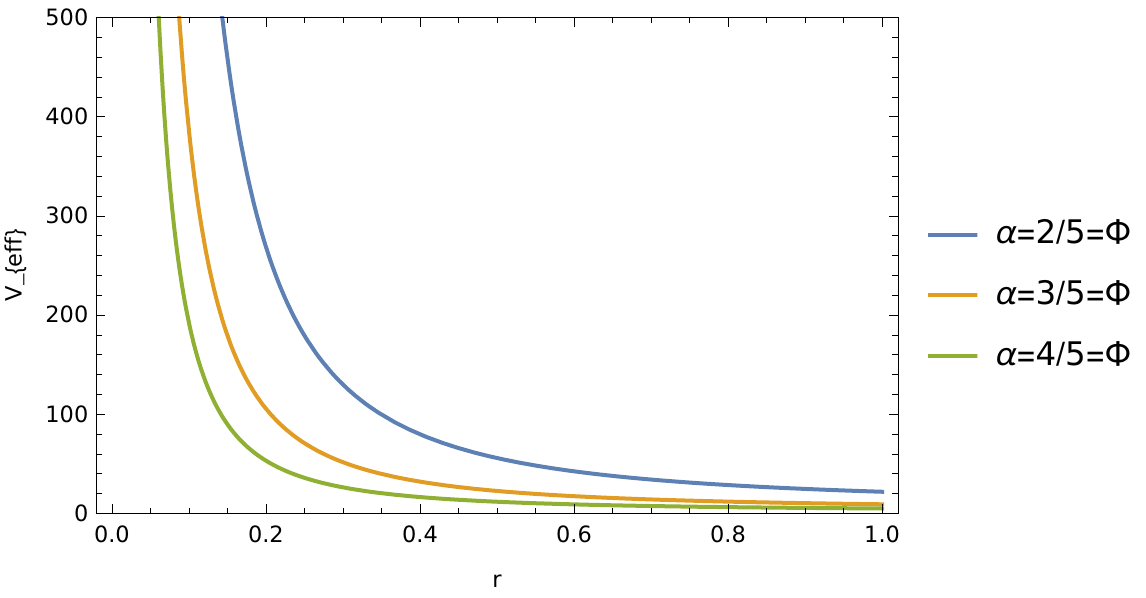}
\caption{$l=1=M=\delta=\gamma=V_0$.}
\label{fig: 4 (c)}
\end{subfigure}
\hfill
\begin{subfigure}[b]{0.4\textwidth}
\includegraphics[width=2.7in, height=1.45in]{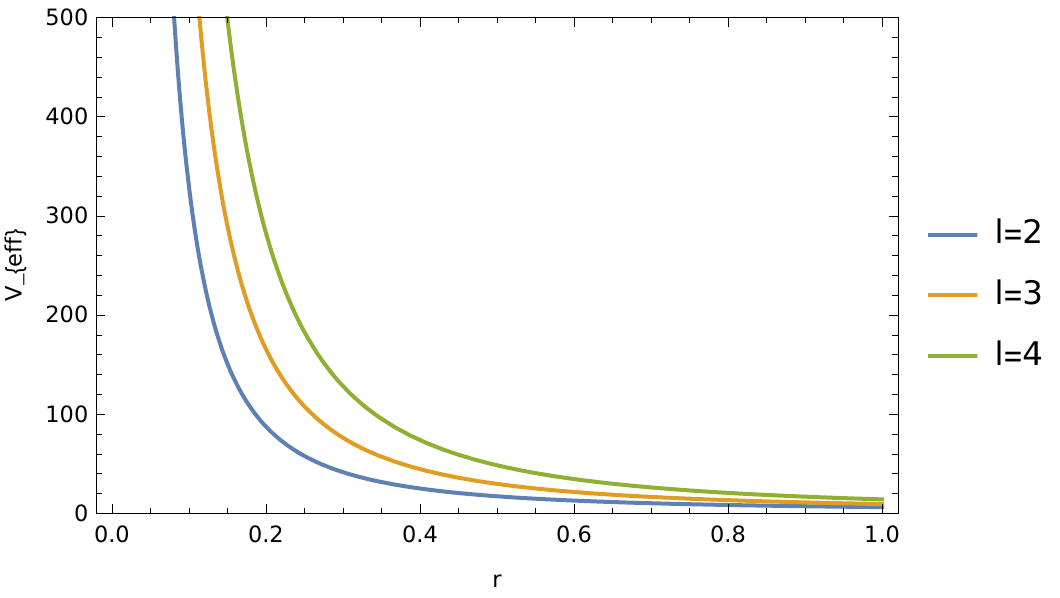}
\caption{$l=1=M=\delta=\gamma=V_0$, $\alpha=4/5$, $\Phi=1$.}
\label{fig: 4 (d)}
\end{subfigure}
\caption{Effective potential of the quantum system using Mie-type potential.}
\label{fig: 4}
\end{figure}

Thereby, substituting the potential (\ref{22}) into the Eq. (\ref{6}), we have obtained the following radial equation:
\begin{equation}
{\rm \psi''(r)+\frac{2}{r}\,\psi'(r)+\Big(-\Delta^2-\frac{j^2}{r^2}-\frac{2\,\kappa}{r}\Big)\,\psi(r)=0},
\label{23}
\end{equation}
where ${\rm j}$ is defined in Eq. (\ref{10}) and
\begin{eqnarray}
{\rm \Delta^2=\frac{2\,M\,(V_0-E)}{\alpha^2}}\quad,\quad {\rm \kappa=\frac{M\,\delta}{\alpha^2}}.
\label{24}
\end{eqnarray}
We perform a change of function via ${\rm \psi (r)=\frac{R(r)}{\sqrt{r}}}$ in the Eq. (\ref{23}), we obtain the following radial wave equation
\begin{equation}
{\rm R''(r)+\frac{1}{r}\,R'(r)+\Big(-\Delta^2-\frac{\tau^2}{r^2}-\frac{2\,\kappa}{r}\Big)\,R(r)=0},
\label{24a}
\end{equation}
where ${\rm \tau=\sqrt{j^2+\frac{1}{4}}=\sqrt{\frac{(l-\Phi)\,(l-\Phi+1)+2\,M\,\gamma}{\alpha^2}+\frac{1}{4}}}=2\,\mu$ (see, Eq. (\ref{14})).

Finally, performing a change of variables via $\xi=2\,\Delta\,r$ in the Eq. (\ref{24a}), we obtain
\begin{equation}
{\rm R''(\xi)+\frac{1}{\xi}\,R'(\xi)+\Bigg(-\frac{\tau^2}{\xi^2}-\frac{\kappa}{\Delta}\,\frac{1}{\xi}-\frac{1}{4}\Bigg)\,R (\xi)=0}.
\label{25}
\end{equation}
It is well-known that the wave function ${\rm R (\xi)}$ is well-behaved at the origin $\xi \to 0$, since it is a singular point of the Eq. (\ref{25}). Suppose, a possible solution to the Eq. (\ref{25}) is given by
\begin{equation}
{\rm R (\xi)=\xi^{\tau}\,e^{-\frac{\xi}{2}}\,F(\xi)}, 
\label{26}
\end{equation}
where ${\rm F (\xi)}$ is an unknown function. 

Thereby, substituting solution (\ref{26}) in the Eq. (\ref{25}), we obtain the following differential equation
\begin{equation}
{\rm \xi\,F''(\xi)+\Big(1+2\,\tau-\xi\Big)\,F'(\xi)+\Bigg(-\tau-\frac{\kappa}{\Delta}-\frac{1}{2}\Bigg)\,F (\xi)=0}.
\label{27}
\end{equation}
Equation (\ref{27}) is the confluent hypergeometric differential equation form \cite{MA,GBA} and ${\rm F(\xi)}$ the confluent hypergeometric function ${\rm F(\xi)={}_1 F_{1} \Big(\tau+\frac{\kappa}{\Delta}+\frac{1}{2}, 2\,\tau+1 ; \xi\Big)}$ which is well-behaved for $\xi \to \infty$. As stated earlier, for the bound-states solution of the quantum system with Mie-type potential, the function ${\rm {}_1 F_{1}}$ is a finite degree polynomial of degree ${\rm n}$, and the quantity ${\rm \Big(\tau+\frac{\kappa}{\Delta}+\frac{1}{2}\Big)=-n}$, where ${\rm n=0,1,2,....}$. 

After simplification of the condition ${\rm \Big(\tau+\frac{\kappa}{\Delta}+\frac{1}{2}\Big)=-n}$ using (\ref{24}), one can obtain the following expression of the energy eigenvalue given by 
\begin{equation}
{\rm E_{n,l}=V_0-\frac{M\,\delta^2}{2\,\Bigg(\Big(n+\frac{1}{2}\Big)\,\alpha+\sqrt{(l-\Phi)\,(l-\Phi+1)+2\,M\,\gamma+\frac{\alpha^2}{4}}\Bigg)^2}}.
\label{28}
\end{equation}

Equation (\ref{28}) is the non-relativistic energy expression of particles confined by the Aharonov-Bohm flux field with Mie-type potential in a point-like global monopole background. We can see that this eigenvalue expression (\ref{28}) is influenced by the topological defects of a point-like global monopole characterized by the parameter $\alpha$, and the quantum flux field ${\rm \Phi_{AB}}$ and gets modified compared to the results obtained in Refs. \cite{SMI,KJO,SMI3,SMI2,SMI4} in the flat space with Mie-type potential. We have plotted graphs of the energy expression ${\rm E_{n,l}}$ (fig. 5) and the radial wave function ${\rm \psi_{n,l}(\xi)}$ (fig. 6) for for different radial mode with various values of parameters. 

\begin{figure}
\begin{subfigure}[b]{0.4\textwidth}
\includegraphics[width=2.8in, height=1.45in]{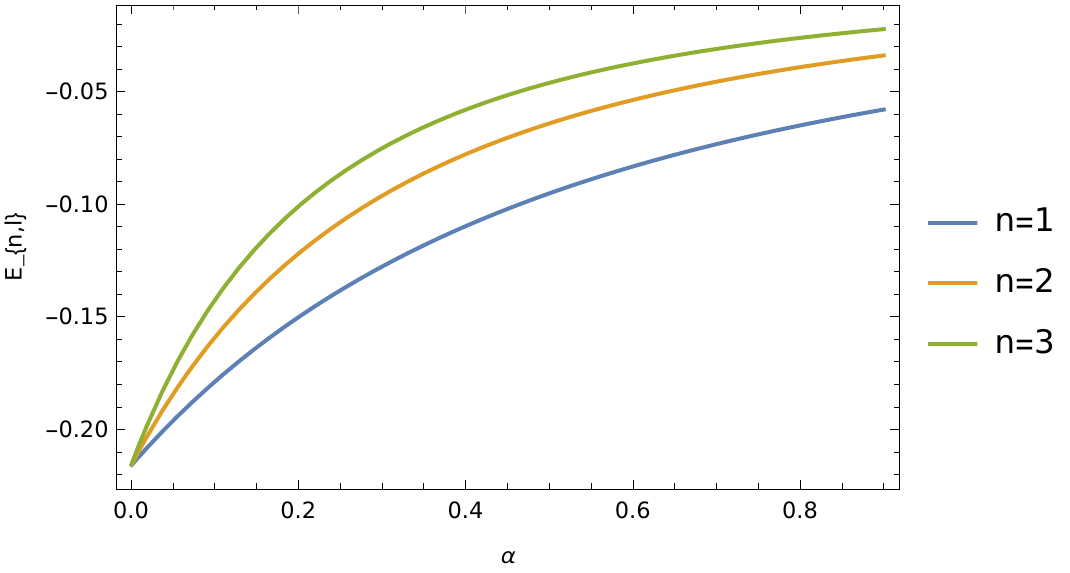}
\caption{$l=1=M=\delta=\gamma=V_0$, $\Phi=3/4$.}
\label{fig: 5 (a)}
\end{subfigure}
\hfill
\begin{subfigure}[b]{0.4\textwidth}
\includegraphics[width=2.95in, height=1.45in]{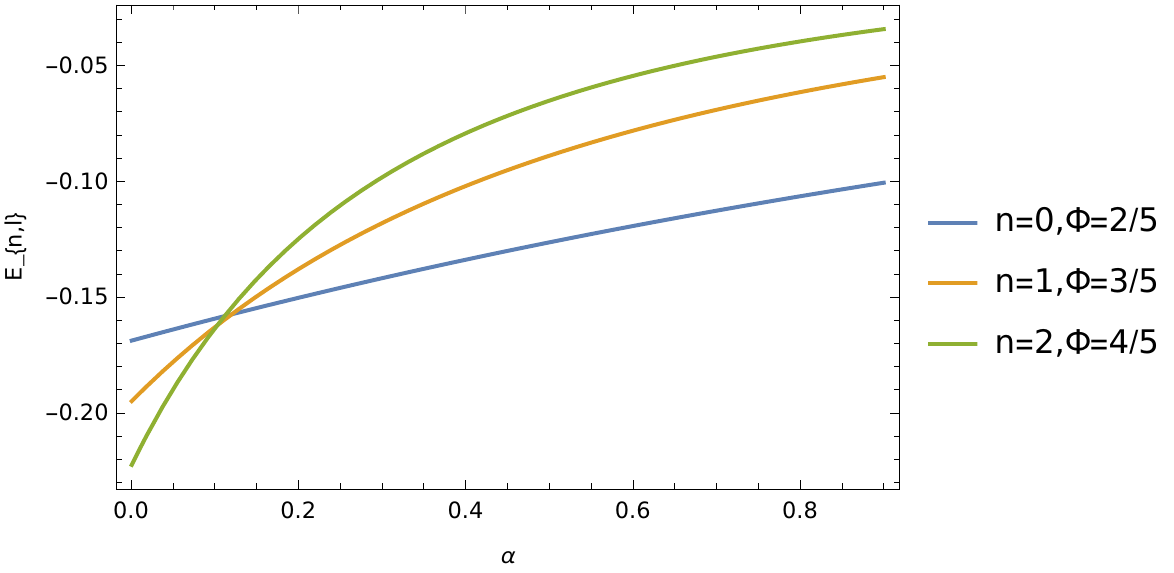}
\caption{$M=1=\delta=\gamma=V_0$.}
\label{fig: 5 (b)}
\end{subfigure}
\hfill\\
\begin{subfigure}[b]{0.4\textwidth}
\includegraphics[width=2.9in, height=1.45in]{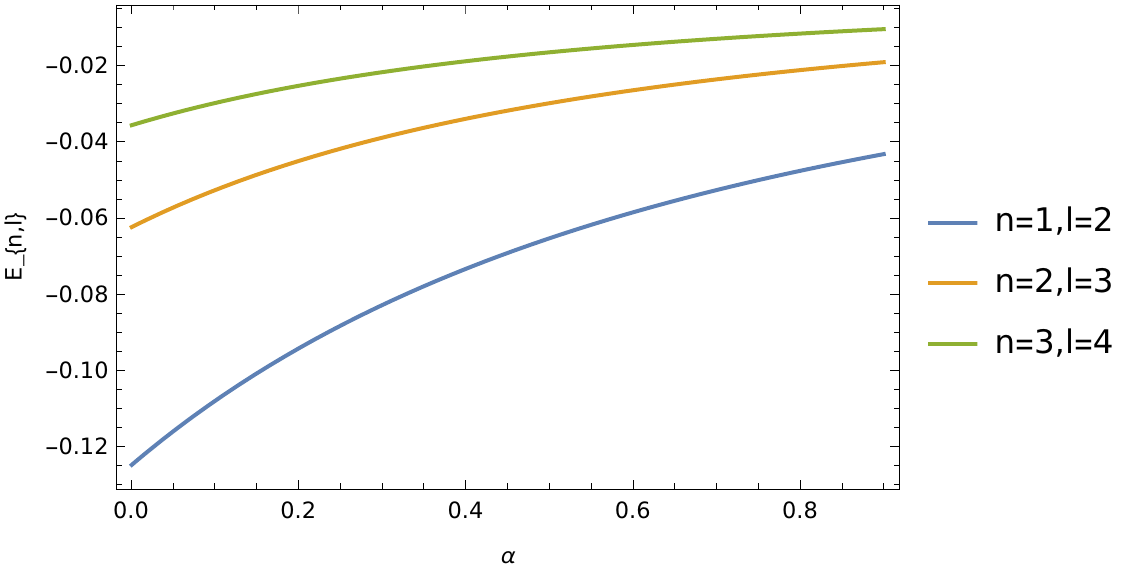}
\caption{$l=1=M=\delta=\gamma=V_0$, $\Phi=1$.}
\label{fig: 5 (c)}
\end{subfigure}
\hfill
\begin{subfigure}[b]{0.4\textwidth}
\includegraphics[width=2.7in, height=1.45in]{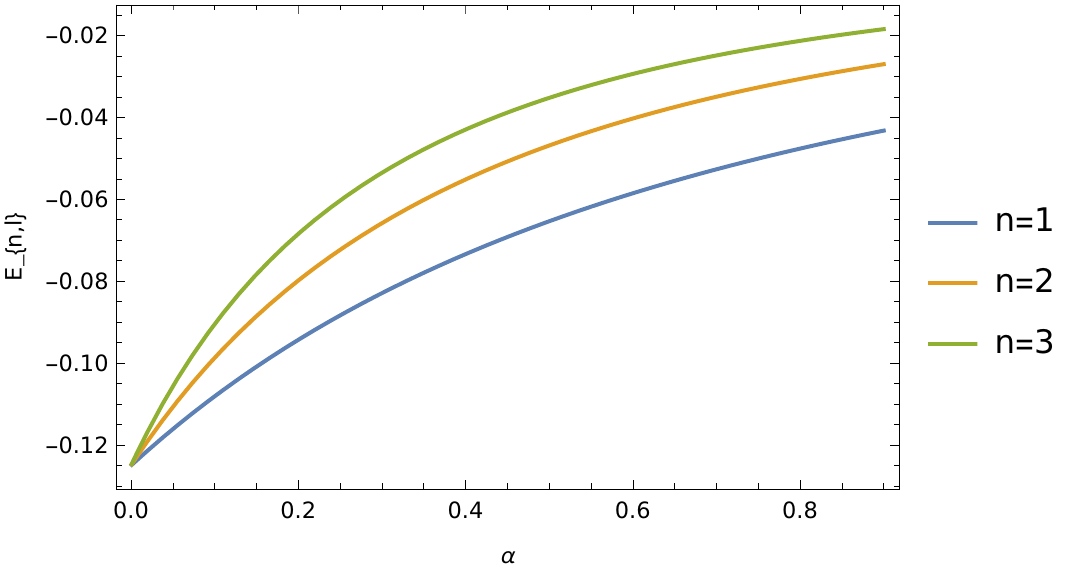}
\caption{$l=1=M=\delta=\gamma=V_0$, $\Phi=0$.}
\label{fig: 5 (d)}
\end{subfigure}
\caption{The energy Eigenvalue $E_{n,l}$ with topological defect parameter $\alpha$ for various values of other parameters.}
\label{fig: 5}
\end{figure}

\begin{figure}
\begin{subfigure}[b]{0.4\textwidth}
\includegraphics[width=2.8in, height=1.45in]{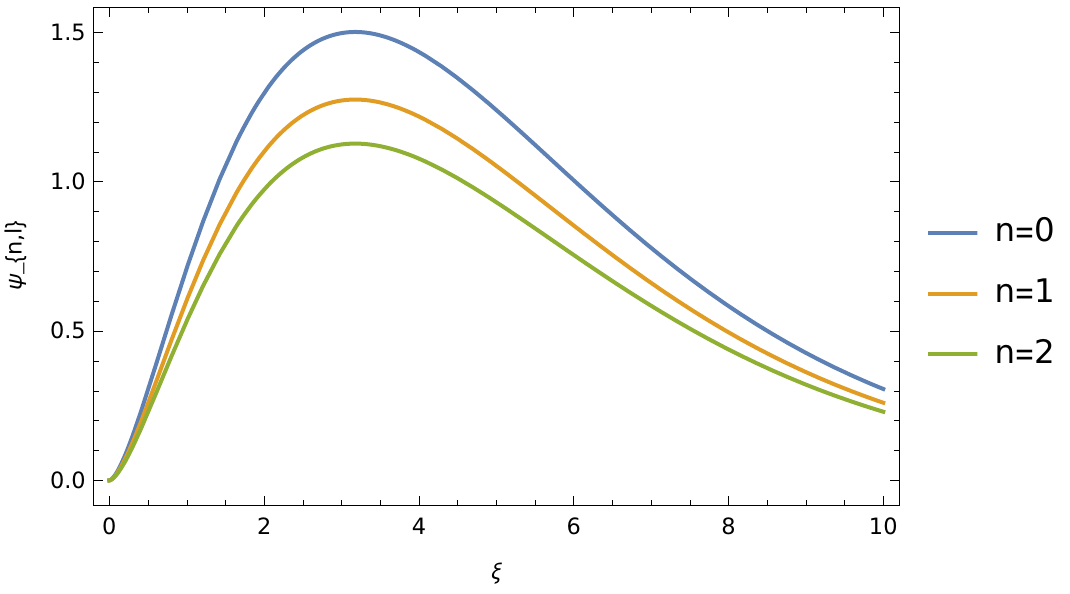}
\caption{$l=1=M=\delta=\gamma$, $\alpha=3/4=\Phi$.}
\label{fig: 6 (a)}
\end{subfigure}
\hfill
\begin{subfigure}[b]{0.4\textwidth}
\includegraphics[width=2.9in, height=1.45in]{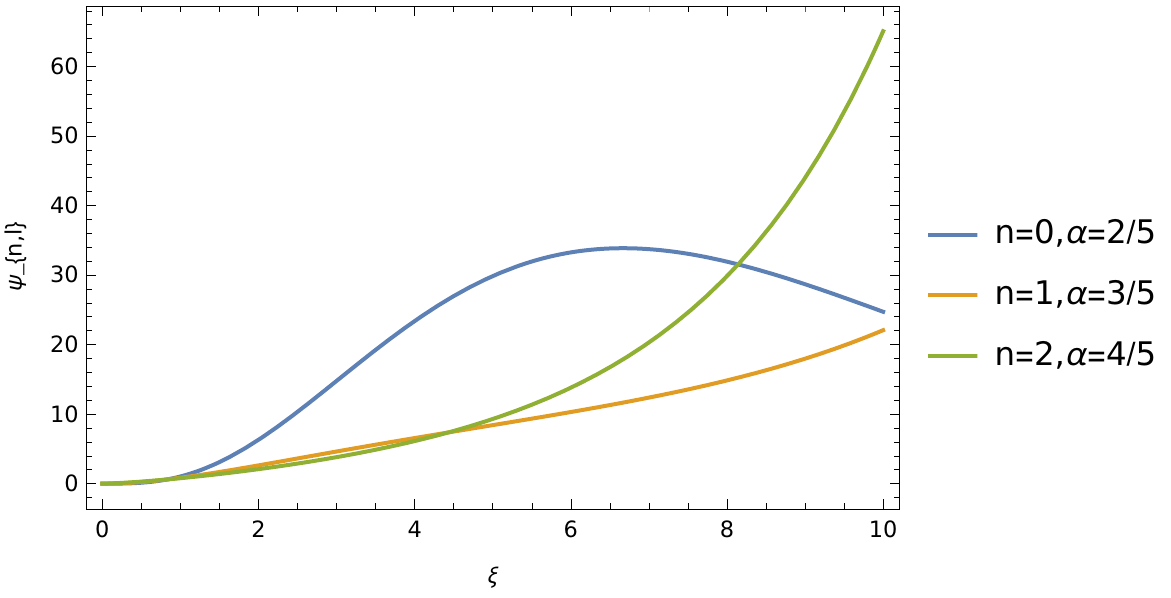}
\caption{$l=1=M=\delta=\gamma$, $\Phi=3/4$.}
\label{fig: 6 (b)}
\end{subfigure}
\hfill\\
\begin{subfigure}[b]{0.4\textwidth}
\includegraphics[width=3.0in, height=1.45in]{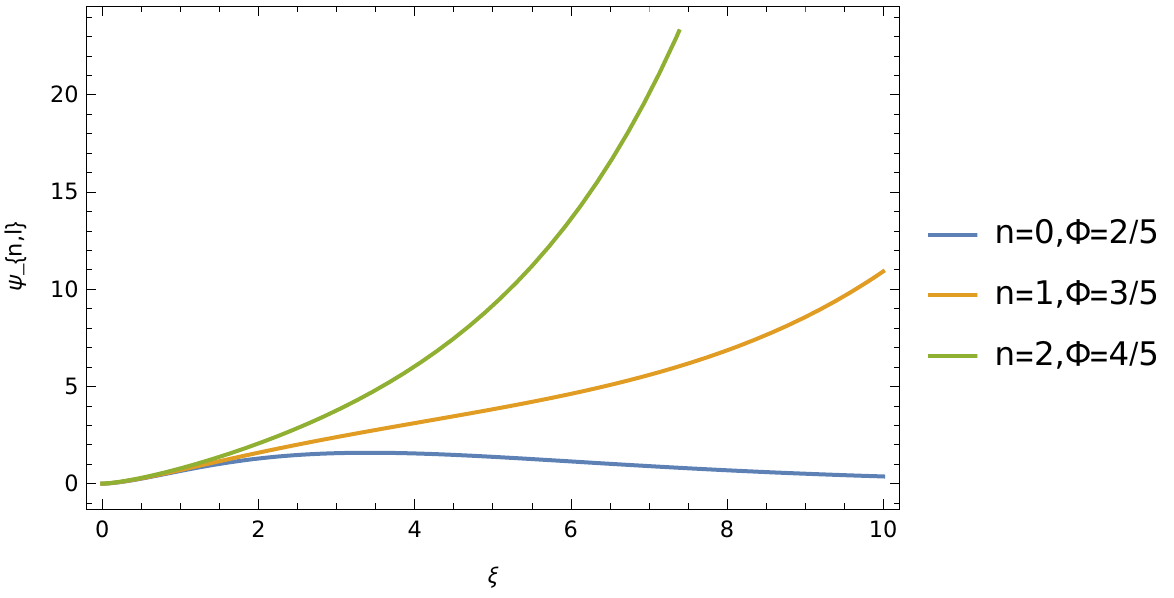}
\caption{$l=1=M=\delta=\gamma$, $\alpha=4/5$.}
\label{fig: 6 (c)}
\end{subfigure}
\hfill
\begin{subfigure}[b]{0.4\textwidth}
\includegraphics[width=2.95in, height=1.45in]{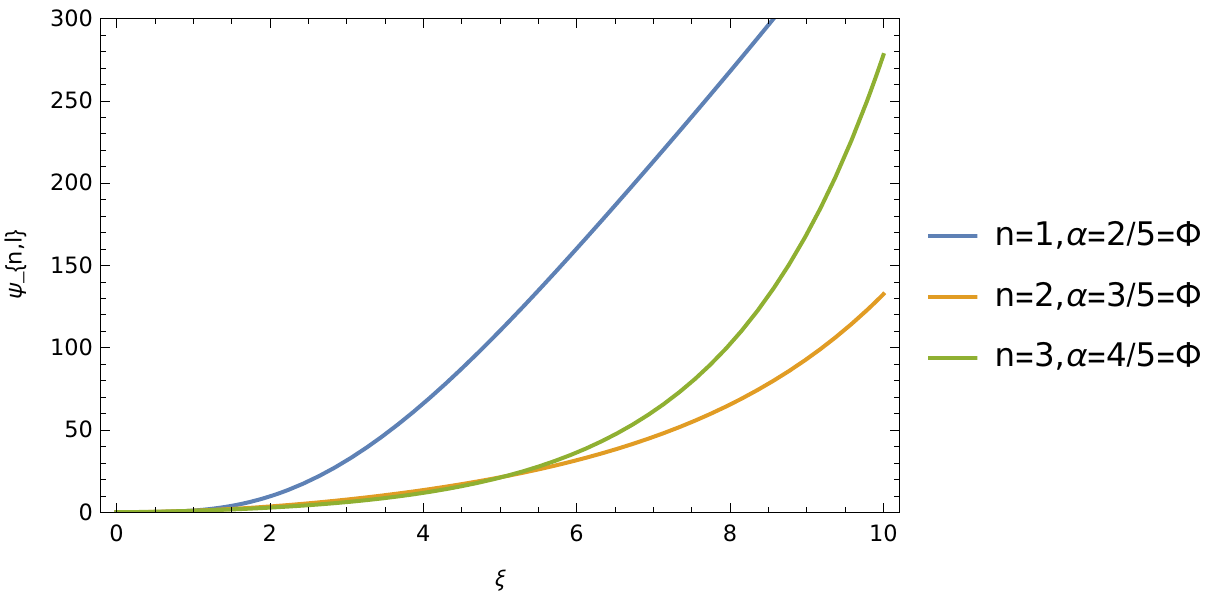}
\caption{$l=1=M=\delta=\gamma$.}
\label{fig: 6 (d)}
\end{subfigure}
\caption{The radial wave function $\psi_{n,l}$ with $\xi$ for various values of different parameters.}
\label{fig: 6}
\end{figure}

The radial wave function is therefore given by
\begin{equation}
{\rm R_{n,l} (\xi)=\xi^{\tau}\,e^{-\frac{\xi}{2}}\,{}_1 F_{1} \Big(-n, 1+2\,\tau ; \xi\Big)}. 
\label{29}
\end{equation}
That can be written as
\begin{eqnarray}
{\rm \psi_{n,l} (\xi)}={\rm \sqrt{2\,\Delta}\,\xi^{\tau-1/2}\,e^{-\frac{\xi}{2}}\,{}_1 F_{1} \Big(-n, 1+2\,\tau ; \xi\Big)}={\rm \sqrt{\frac{2\,M\,\delta}{\alpha^2\,\Big(n+\tau+\frac{1}{2}\Big)}}\,\xi^{\tau-1/2}\,e^{-\frac{\xi}{2}}\,{}_1 F_{1} \Big(-n, 1+2\,\tau ; \xi\Big)}.
\label{30}
\end{eqnarray}

If we analyze the quantum system without the topological defect, that is, $\alpha \to 1$, the space-time geometry (\ref{1}) under this becomes Minkowski flat space. Therefore, for $\alpha \to 1$, the energy eigenvalue expression from (\ref{28}) becomes
\begin{equation}
{\rm E_{n,l}=V_0-\frac{M\,\delta^2}{2\,\Bigg(n+\sqrt{\Big(l-\Phi+\frac{1}{2}\Big)^2+2\,M\,\gamma}+\frac{1}{2}\Bigg)^2}}
\label{31}
\end{equation}
which is similar to the result obtained in Refs. \cite{bb40,RS} provided zero magnetic flux ${\rm \Phi \to 0}$ here. Thus, we can see that the presence of the quantum flux field modified the energy spectrum of non-relativistic particles compared to those results obtained in Refs. \cite{bb40,RS}. Furthermore, the presence of the topological defect characterized by the parameter $\alpha$ modified the eigenvalue expression (\ref{28}) more in addition to the quantum flux compared to those results obtained in Refs. \cite{bb40,RS}. 

\section{Applications to Some Diatomic Molecular potentials}

The above results of the studied quantum system previous are now being utilized to develop solutions to some specific types of physical potential models which have application in practical problems. The pseudo-harmonic and Kratzer potentials are successfully employed \cite{GL} to study the eigenvalues spectra of a class of diatomic molecules in flat space background. Here, we will study the quantum system in curved space-time background under topological defects with these known potentials.

\subsection{\bf Harmonic Oscillator Potential}

The harmonic oscillator potential can be recovered from the potential (\ref{9}) by setting the potential parameters $\gamma=0$, $V_0=0$, and ${\rm \beta=\frac{1}{2}\,M\,\omega^2_0}$, we have, ${\rm V (r)=\frac{1}{2}\,M\,\omega^2_0\,r^2}$, a harmonic oscillator potential. Here ${\rm \omega_0}$ is the oscillator frequency. Therefore, using harmonic oscillator potential in the quantum system, one will find the following energy eigenvalue expression:
\begin{eqnarray}
{\rm E_{n,l}=2\,\alpha\,\omega_0\,\Bigg(n+\frac{1}{2\,\alpha}\,\sqrt{(l-\Phi)\,(l-\Phi+1)+\frac{\alpha^2}{4}}+\frac{1}{2}\Bigg)}.
\label{cc1}
\end{eqnarray}
Equation (\ref{cc1}) is the energy levels of harmonic oscillator in the presence of the Aharonov-Bohm flux field in a point-like global monopole. 

In absence of quantum flux field, ${\rm \Phi_{AB} \to 0}$, the energy eigenvalues from Eq. (\ref{cc1}) becomes
\begin{eqnarray}
{\rm E_{n,l}=\alpha\,\omega_0\,\Bigg(2\,n+\sqrt{\frac{l\,(l+1)}{\alpha^2}+\frac{1}{4}}+1\Bigg)}.
\label{cc2}
\end{eqnarray}
which is similar to those results obtained in Refs. \cite{CF,CF2}. Thus, we can see that the quantum flux field shifts the eigenvalue solution of harmonic oscillator compared to those results obtained in Refs. \cite{CF,CF2}. The energy eigenvalue ${\rm E_{n,l}}$ depends on the geometric quantum phase ${\rm \Phi_B}$ that shows an analogue of the Aharonov-Bohm effect \cite{YA,MP}.

\subsection{\bf Pseudoharmonic Potential}

The pseudoharmonic potential can be recovered from the potential (\ref{9}) by setting the parameters ${\rm \beta=\frac{D_e}{r^{2}_{0}}}$, ${\rm \gamma=D_e\,r^{2}_{0}}$, and ${\rm V_0=-2\,D_e}$. Thus, we obtain the following potential form \cite{SMI,RSS,FC2}
\begin{equation}
{\rm V(r)=\frac{D_e}{r^{2}_{0}}\,r^2+\frac{D_e\,r^{2}_{0}}{r^2}-2\,D_e=D_e\,\Bigg(\frac{r}{r_0}-\frac{r_0}{r}\Bigg)^2}.
\label{cc3}
\end{equation}
Here ${\rm D_e}$ is the dissociation energy between atoms in a solid and ${\rm r_0}$ is the equilibrium inter-nuclear separation. This potential is of great importance not only in physics but also in chemistry and is useful to describe the interactions of some diatomic molecules \cite{AC,SMI5}.

Thereby, using pseudoharmonic potential in the quantum and solving the wave equation, one can find the energy eigenvalue expression as follows
\begin{eqnarray}
{\rm E_{n,l}=-2\,D_e+\alpha\,\sqrt{\frac{2\,D_e}{M\,r^2_{0}}}\,\Bigg(2\,n+\sqrt{\frac{(l-\Phi)\,(l-\Phi+1)+2\,M\,D_e\,r^2_{0}}{\alpha^2}+\frac{1}{4}}+1\Bigg)}.
\label{cc4}
\end{eqnarray}
The normalized radial wave functions are given by
\begin{equation}
{\rm \psi_{n,l} (s)=B_{n,l}\,\Bigg(\frac{2\,M\,D_e}{\alpha^2\,r^2_{0}}\Bigg)^{3/8}\,s^{\mu-\frac{1}{4}}\,e^{-\frac{s}{2}}\,{}_1 F_{1} (-n,1+2\,\mu;s)},\quad {\rm B_{n,l}=\frac{1}{\left(\sqrt{j^2+\frac{1}{4}}\right)!}\,\sqrt{\frac{2\,\alpha\,\left(n+\sqrt{j^2+\frac{1}{4}}\right)!}{n!}}},
\label{cc5}
\end{equation}
where ${\rm \tau=2\,\mu=\sqrt{j^2+1/4}}$, and ${\rm j=\sqrt{\frac{(l-\Phi)\,(l-\Phi+1)+2\,M\,D_e\,r^2_{0}}{\alpha^2}}}$.

Equation (\ref{cc4}) is the energy eigenvalue expression and Eq. (\ref{cc5}) is the normalized radial wave function of non-relativistic particles confined by the quantum flux field with pseudoharmonic potential in a point-like global monopole background. One can see that the eigenvalue solution Eqs. (\ref{cc4})--(\ref{cc5}) get modified by the topological defect characterized by the parameter $\alpha$, and the quantum flux field compared to those results obtained in Refs. \cite{SMI,RSS,FC2}.

For zero quantum flux field, ${\rm \Phi_{AB} \to 0}$, the energy eigenvalue of the non-relativistic particles from Eq. (\ref{cc4}) becomes
\begin{eqnarray}
{\rm E_{n,l}=-2\,D_e+\alpha\,\sqrt{\frac{2\,D_e}{M\,r^2_{0}}}\,\Bigg(2\,n+\sqrt{\frac{l\,(l+1)+2\,M\,D_e\,r^2_{0}}{\alpha^2}+\frac{1}{4}}+1\Bigg)}.
\label{cc6}
\end{eqnarray}
And that the corresponding normalized wave function will be
\begin{eqnarray}
\psi_{n,l} (s)&=&{\rm B_{n,l}\,\Bigg(\frac{2\,M\,D_e}{\alpha^2\,r^2_{0}}\Bigg)^{3/8}\,s^{\Big[-\frac{1}{4}+\frac{1}{2}\,\sqrt{\frac{l\,(l+1)+2\,M\,D_e\,r^2_{0}}{\alpha^2}+\frac{1}{4}}\Big]}\,e^{-\frac{s}{2}}\,{}_1 F_{1} (-n,1+2\,\mu;s)},\nonumber\\
{\rm B_{n,l}}&=&{\rm \frac{1}{\left(\sqrt{\frac{l\,(l+1)+2\,M\,D_e\,r^2_{0}}{\alpha^2}+\frac{1}{4}}\right)!}\,\sqrt{\frac{2\,\alpha\,\left(n+\sqrt{\frac{l\,(l+1)+2\,M\,D_e\,r^2_{0}}{\alpha^2}+\frac{1}{4}}\right)!}{n!}}}
\label{ccc6}
\end{eqnarray}
We can see that the global effects of the geometry characterized by the parameter $\alpha$ of a point-like global monopole modified the result compared to the flat space with pseudoharmonic potential obtained in Refs. \cite{SMI,RSS,FC2}.

\subsection{\bf The Kratzer-Fues or Modified Kratzer Potential}

The modified Kratzer potential can easily be recovered from the potential (\ref{22}) by setting the potential parameters ${\rm \delta=-2\,D_e\,r_{0}}$, ${\rm \gamma=D_e\,r^{2}_{0}}$, and ${\rm V_0=D_e}$, we have \cite{SMI6}
\begin{equation}
{\rm V(r)=D_e+\frac{D_e\,r^{2}_{0}}{r^{2}}-\frac{2\,D_e\,r_{0}}{r}=D_e\,\Bigg(\frac{r-r_0}{r}\Bigg)^2}.
\label{cc7}
\end{equation}
This potential also known as Kratzer-Fues potential and has been used by several authors \cite{KJO,SMI2,SMI4,RSS,FC2,VK} to describe the molecular structures and interactions. 

Thereby, using this potential in the quantum system and solving the wave equation, one will find the energy eigenvalue expression as follows 
\begin{equation}
{\rm E_{n,l}=D_e-\frac{2\,M\,D^2_{e}\,r^2_{0}}{\Bigg(n\,\alpha+\frac{\alpha}{2}+\sqrt{(l-\Phi)\,(l-\Phi+1)+2\,M\,D_e\,r^2_{0}+\frac{\alpha^2}{4}}\Bigg)^2}}.
\label{cc8}
\end{equation}
The radial wave function is therefore given by
\begin{equation}
{\rm R_{n,l} (\xi)=\xi^{\tau}\,e^{-\frac{\xi}{2}}\,{}_1 F_{1} \Big(-n, 1+2\,\tau ; \xi\Big)}, 
\label{cc9}
\end{equation}
where ${\rm \tau=\sqrt{\frac{(l-\Phi)\,(l-\Phi+1)+2\,M\,D_e\,r^2_{0}}{\alpha^2}+\frac{1}{4}}}$.

Equation (\ref{cc8}) is the non-relativistic energy expression and Eq. (\ref{cc9}) is the radial wave function of non-relativistic particles confined by the quantum flux field with modified Kratzer potential in a point-like defect. One can see that the eigenvalue solution gets modified by the topological defect of the geometry characterized by the parameter $\alpha$, and the flux field ${\rm \Phi_{AB}}$ compared to those results obtained in Refs. \cite{SMI,KJO,SMI3,SMI2,RSS,FC2,SMI4,SMI6,VK} with this potential.

If we analyze the quantum system without the topological defect, that is, $\alpha \to 1$, the space-time geometry becomes flat space. Therefore, $\alpha \to 1$. the energy eigenvalue expression from Eq. (\ref{cc8}) becomes
\begin{equation}
{\rm E_{n,l}=D_e-\frac{2\,M\,D^2_{e}\,r^2_{0}}{\Bigg(n+\frac{1}{2}+\sqrt{(l-\Phi)\,(l-\Phi+1)+2\,M\,D_e\,r^2_{0}+\frac{1}{4}}\Bigg)^2}}.
\label{cc10}
\end{equation}
The radial wave functions are given by
\begin{equation}
{\rm R_{n,l} (\xi)=\xi^{\tau_0}\,e^{-\frac{\xi}{2}}\,{}_1 F_{1} \Big(-n, 1+2\,\tau_0 ; \xi\Big)}, 
\label{cc11}
\end{equation}
where ${\rm \tau_0=\sqrt{(l-\Phi)\,(l-\Phi+1)+2\,M\,D_e\,r^2_{0}+\frac{1}{4}}}$.

Equation (\ref{cc10}) is the energy spectra and Eq. (\ref{cc11}) is the radial wave function of non-relativistic particles confined by the quantum flux field with modified Kratzer potential in flat space background.

\subsection{\bf Kratzer Potential}

The Kratzer potential can be recovered from potential (\ref{22}) by setting the potential parameters ${\rm \delta=-2\,D_e\,r_0}$, ${\rm \gamma=D_e\,r^2_{0}}$, and $V_0=0$, we obtain \cite{AK,EF,HA,FC2,gg1,gg2,gg4}
\begin{equation}
{\rm V(r)=\frac{D_e\,r^2_{0}}{r^{2}}-\frac{2\,D_e\,r_0}{r}=2\,D_e\,\Big[\frac{1}{2}\,\Big(\frac{r_0}{r}\Big)^2-\Big(\frac{r_0}{r}\Big)\Big]}.
\label{cc12}
\end{equation}
This potential has extensively been used to describe the molecular structures and interactions in chemistry. 

Thereby, using Kratzer potential in the quantum system and solving the wave equation, one can obtain bound-states energy eigenvalue expression as follows 
\begin{equation}
{\rm E_{n,l}=-\frac{2\,M\,D^2_{e}\,r^2_{0}}{\Bigg(\Big(n+\frac{1}{2}\Big)\,\alpha+\sqrt{(l-\Phi)\,(l-\Phi+1)+2\,M\,D_e\,r^2_{0}+\frac{\alpha^2}{4}}\Bigg)^2}}.
\label{cc13}
\end{equation}
The radial wave functions are given by
\begin{equation}
{\rm R_{n,l} (\xi)=\xi^{\tau}\,e^{-\frac{\xi}{2}}\,{}_1 F_{1} \Big(-n, 1+2\,\tau ; \xi\Big)}, 
\label{cc14}
\end{equation}
where ${\rm \tau=\sqrt{\frac{(l-\Phi)\,(l-\Phi+1)+2\,M\,D_e\,r^2_{0}}{\alpha^2}+\frac{1}{4}}}$.

Equation (\ref{cc13}) is the energy expression and Eq. (\ref{cc14}) is the radial wave function of non-relativistic particles confined by the quantum flux field with Kratzer potential in a point-like defect. One can see that the eigenvalue solution are influenced by the topological defect characterized by the parameter $\alpha$, and the quantum flux field ${\rm \Phi_{AB}}$ and gets them modified compared to those results obtained in Refs. \cite{HA,FC2,gg1,gg2,gg4}.

\section{Conclusions}

In this analysis, we have shown that in the presence of different kinds of interaction potential models, the topological defect changes the physical properties of the quantum system under investigation. Furthermore, the non-relativistic particles under the influences of the quantum flux field but experiencing no direct interactions with a magnetic field also modified the properties of the system. Throughout the analysis, we have observed a shifting in the orbital quantum number ${\rm l}$, that is, ${\rm l \to l'=\left(l-\frac{e\,\Phi_B}{2\pi}\right)}$ called an effective orbital quantum number that depends on the flux field. This effective quantum number appeared in the energy expressions and thus, the eigenvalue depends on the geometric quantum phase $\Phi_B$ and is a periodic function with a periodicity $\Phi_0$, that is, ${\rm E_{n,l} (\Phi_B \pm\,\Phi_0\,\iota)=E_{n,l\mp\iota} (\Phi_B)}$, where $\iota=0,1,2,...$. This dependence of the eigenvalue on the geometric phase shows an analogue of the Aharonov-Bohm (AB) effect \cite{YA,MP} for the bound-states. It is well-known in condensed matter physics that the dependence of the eigenvalue on geometric phase gives rise to a persistent currents which we will discuss in the future work.

We derived the radial wave equation of the Schr\"{o}dinger equation with the quantum flux field and potential in a topological defect space-time produced by a point-like global monopole. Then, in {\tt sub-section 2.1}, we have considered pseudoharmonic-type potential and solved the radial equation analytically. The exact eigenvalue expression of the particles are given by the Eq. (\ref{16}) and the normalized radial wave function by Eq. (\ref{17})--(\ref{20}). We have shown that the topological defect characterized by the parameter $\alpha$ and the quantum flux field influences the eigenvalue solution and gets them modified compared to those results obtained in Refs. \cite{SMI,SHD,KJO,SMI3,VK} in the flat space. As special cases, we have analyzed the eigenvalue solution without topological effects, and have shown that only the flux field shifted the energy levels and wave function compared to those results obtained in Refs. \cite{FC,bb40}. 

In {\tt sub-section 2.2}, we have considered the Mie-type potential and solved the quantum mechanical problem in the background of a point-like defect analytically. The energy eigenvalues expression are given by the Eq. (\ref{28}) and radial wave function by (\ref{30}) of the particles. Here also, we have shown that the topological defect of a point-like defect characterized by the parameter $\alpha$, and the quantum flux field influences the eigenvalue solution and modified the result compared to the flat space results obtained in Refs. \cite{SMI,KJO,SMI3,SMI2,SMI4}. Also, the presented eigenvalue solution is analyzed without topological effects and have shown that the quantum flux field shifted the result compared to those obtained in the flat space in Refs. \cite{bb40,RS}.

In {\tt section 3}, we have utilized these eigenvalue solutions of the quantum system obtained in {\tt sub-section 2.1} and {\tt sub-section 2.2} for some known interaction potential models which have been used for diatomic molecular structure in physics and chemistry. These potential models include harmonic oscillator potential ({\tt sub-section 3.1}), pseudoharmonic potential ({\tt sub-section 3.2}), modified Kratzer or Kratzer-Fues potential ({\tt sub-section 3.3}), and Kratzer potential ({\tt sub-section 3.4}). With these potential models, we have presented the energy eigenvalue expressions and radial wave functions of the non-relativistic particles under the influences of the quantum flux field in a point-like global monopole. We have shown that both the topological defect of a point-like global monopole space-time characterized by the parameter $\alpha$ and the quantum flux field $\Phi_{AB}$ modified the eigenvalue solutions compared to the known results obtained in the flat space in the literature.

\section*{Acknowledgement}

We sincerely acknowledged the anonymous kind referee(s) for valuable comments and suggestions.

\section*{Conflict of Interest}

There is no conflict of interests regarding publication of this paper.

\section*{Funding Statement}

No fund has received for this paper.

\section*{Data Availability Statement}

No new data are generated in this paper.

\section*{Appendix A : The solutions of angular equations}

\setcounter{equation}{0}
\renewcommand{\theequation}{A.\arabic{equation}}

Expressing the wave equation (\ref{2}) in the space-time background (\ref{1}) and using Eqs. (\ref{3})-(\ref{4}), we obtain the following equation
\begin{eqnarray}
&&\frac{1}{\psi(r)}\Bigg[\alpha^2\,\frac{d\psi(r)}{dr}\,\Big(r^2\,\frac{d\psi(r)}{dr}\Big)+2\,M\,(E-V(r))\,r^2\,\psi(r)\Bigg]\nonumber\\
&&=-\frac{1}{Y(\theta,\phi)}\Bigg[\frac{1}{\sin \theta}\,\frac{d}{d\theta}\,\Big(\sin \theta\,\frac{d\,Y (\theta,\phi)}{d\theta}\Big)+\frac{1}{\sin^2 \theta}\,\Big(\frac{d}{d\phi}-i\,\Phi\Big)^2\,Y (\theta,\phi)\Bigg].
\label{A.1}
\end{eqnarray}

Writing both side equal to $l'(l'+1)$, we obtain the radial equation (\ref{5}). The angular equation is given by 
\begin{equation}
{\rm \frac{1}{\sin \theta}\,\frac{d}{d\theta}\,\Big(\sin \theta\,\frac{d\,Y (\theta,\phi)}{d\theta}\Big)+\frac{1}{\sin^2 \theta}\,\Big(\frac{d}{d\phi}-i\,\Phi\Big)^2\,Y (\theta,\phi)=-l'\,(l'+1)\,Y(\theta,\phi)},
\label{A.2}
\end{equation}
where ${\rm l'=(l-\Phi)}$ (will explain later on).

The above equation (\ref{A.2}) can be written as
\begin{equation}
{\rm \sin \theta\,\frac{d}{d\theta}\,\Big(\sin \theta\,\frac{d\,Y (\theta,\phi)}{d\theta}\Big)+l'\,(l'+1)\,\sin^2\,\theta\,Y(\theta,\phi)+\Big(\frac{d}{d\phi}-i\,\Phi\Big)^2\,Y (\theta,\phi)}=0.
\label{A.3}
\end{equation}

Assuming a separable solution to the Eq (\ref{A.3}), substitution of ${\rm Y (\theta,\phi)=A(\theta)\,B(\phi)}$ yields 
\begin{equation}
{\rm \frac{1}{A(\theta)}\,\Bigg[\sin \theta\,\frac{d}{d\theta}\,\Big(\sin \theta\,\frac{d\,A (\theta)}{d\theta}\Big)\Bigg]+l'\,(l'+1)\,\sin^2\,\theta}=-{\rm \frac{1}{B(\phi)}\,\Bigg[\Big(\frac{d}{d\phi}-i\,\Phi\Big)^2\,B (\phi)\Bigg]}.
\label{A.4}
\end{equation}

Considering a separation constant equal to $b^2$, we have the azimuthal equation
\begin{eqnarray}
{\rm -\frac{1}{B(\phi)}\,\Big[\Big(\frac{d}{d\phi}-i\,\Phi\Big)^2\,B (\phi)\Big]=b^2\Rightarrow 
\Big(\frac{d}{d\phi}-i\,\Phi\Big)^2\,B(\phi)+b^2\,B(\phi)}=0.
\label{A.5}
\end{eqnarray}

Suppose, a possible solution to the Eq. (\ref{A.5}) is given by
\begin{equation}
{\rm B(\phi)=\frac{1}{\sqrt{2\,\pi}}\,e^{i\,m\,\phi}}.
\label{A.6}
\end{equation}

Substituting the solution (\ref{A.6}) in Eq. (\ref{A.5}), we obtain
\begin{equation}
{\rm (-m^2+2\,m\,\Phi-\Phi^2)\,B(\phi)+b^2\,B(\phi)}=0\Rightarrow {\rm b^2=(m-\Phi)^2=m'^2},
\label{A.7}
\end{equation}
where ${\rm m'=(m-\Phi)}$ and $m$ is the magnetic quantum number.

And the polar $A(\theta)$ equation is
\begin{eqnarray}
&&{\rm \frac{1}{A(\theta)}\,\Bigg[\sin \theta\,\frac{d}{d\theta}\,\Big(\sin \theta\,\frac{d\,A (\theta)}{d\theta}\Big)\Bigg]+l'\,(l'+1)\,\sin^2\,\theta}=m'^2\nonumber\\\Rightarrow
&&{\rm \frac{1}{\sin \theta}\,\frac{d}{d\theta}\,\Big(\sin \theta\,\frac{d\,A(\theta)}{d\theta}\Big) +\Bigg[l'\,(l'+1)-\frac{m'^2}{\sin^2 \theta}\Bigg]\,A(\theta)}=0.
\label{A.8}
\end{eqnarray}

Let us change a variable by ${\rm x=\cos \theta}$. The equation with the function $A(\theta)$ becomes
\begin{equation}
{\rm \frac{d}{dx}\,\Big[(1-x^2)\,\frac{dA(x)}{dx}\Big]+\Big[l'\,(l'+1)-\frac{m'^2}{1-x^2}\Big]\,A(x)}=0.
\label{A.9}
\end{equation}
which is the associated Legendre polynomials equation whose solutions are given by
\begin{equation}
{\rm A(x)=P^{m'}_{l'}(x)=(-1)^{m'}\,(1-x^2)^{m'/2}\,\frac{d^{m'}}{dx^{m'}}\,(P_{l'}(x))},
\label{A.10}
\end{equation}
where superscript $m'$ indicates the order and ${\rm P_{l'} (x)}$ are polynomials of degree $l'$ given by 
\begin{equation}
{\rm P_{l'} (x)=\frac{1}{2^{l'}\,l'!}\,\frac{d^{l'}}{dx^{l'}}\,(x^2-1)^{l'}}.
\label{A.11}
\end{equation}
Noted that the Legendre polynomials ${\rm P_{l'} (x)}$ are polynomials of order $l'$ provided the magnitude of $m'$ must have values less than or equal to $l'$. That is
\begin{equation}
{\rm l'\geq |m'|}\quad \mbox{or}\quad {\rm l'=|m'|+\kappa},\quad {\rm \kappa=0,1,2,....}.
\label{A.12}
\end{equation}
Throughout the analysis we have written ${\rm m'=(m-\Phi)>0}$ always positive, and hence, ${\rm l'=(l-\Phi)}$ due to the quantum flux field $\Phi_{AB}$ present in the quantum system. For zero magnetic flux $\Phi \to 0$, one will get back the standard azimuthal and polar equations which were given in many textbooks. In that case, the relation (\ref{A.12}) can be written as ${\rm l=(|m|+\kappa})$. The shifting in the quantum numbers $l \to l'$ and $m \to m'$ are respectively called the effective orbital and magnetic quantum numbers of the quantum syste.

\end{document}